\documentclass[preprint,11pt]{elsarticle}

\usepackage{amsmath}
\usepackage{amssymb}
\usepackage{amsfonts}
\usepackage{booktabs}
\usepackage{enumitem}
\usepackage{fancyhdr}
\usepackage[a4paper]{geometry}
\usepackage{listings}
\usepackage{placeins}
\usepackage{siunitx}
\usepackage{tikz}
\usetikzlibrary{shapes,arrows, matrix, decorations.pathreplacing}

\usepackage{notoccite}

\usepackage[lowercase]{theoremref}
\usepackage{tikz-uml}
\usepackage{graphicx}
\usepackage{caption}
\usepackage{subcaption}
\usepackage{verbatim}
\usepackage{verbatimbox}
\usepackage{xcolor, colortbl}
\usepackage[hidelinks]{hyperref}
\usepackage{pdfpages}
\usepackage{physics}
\usepackage[titletoc,title]{appendix}
\usepackage{booktabs}
\usepackage[capitalise]{cleveref}
\usepackage{float}

\usepackage{etoolbox}
\usepackage{listings}

\usepackage{titlesec}
\usepackage{tocloft}

\usepackage{apptools}

\usepackage{algorithm}
\usepackage{algpseudocode}

\usetikzlibrary{shapes, arrows.meta, positioning}

\begin{document}

\begin{frontmatter}



\title{A Massively Parallel Performance Portable Free-space Spectral Poisson Solver}


\author[inst1,inst2]{Sonali Mayani}
\author[inst2]{Veronica Montanaro}
\author[inst3,inst4]{Antoine Cerfon}
\author[inst5]{Matthias Frey}

\author[inst6]{Sriramkrishnan Muralikrishnan}
\author[inst1,inst2]{Andreas Adelmann}

\affiliation[inst1]{organization={Paul Scherrer Institute, Villigen},
            country={Switzerland}}
\affiliation[inst2]{organization={ETH Zürich},
            country={Switzerland}}
\affiliation[inst3]{organization={Courant Institute of Mathematical Sciences, New York University},
            country={USA}}
\affiliation[inst4]{organization={Type One Energy Group - Canada Inc},
            addressline={Vancouver, BC},
            country={Canada}}
\affiliation[inst5]{organization={Mathematical Institute},
            addressline={University of St Andrews},
            postcode={KY16 9SS},
            country={UK}}
\affiliation[inst6]{organization={Jülich Supercomputing Centre, Forschungszentrum Jülich GmbH},
            country={Germany}}

\begin{abstract}

Vico et al. (2016) suggest a fast algorithm for computing volume potentials, beneficial to fields with problems requiring the solution of the free-space Poisson’s equation, such as beam and plasma physics. Currently, the standard is the algorithm of Hockney and Eastwood (1988), with second order in convergence at best. The algorithm proposed by Vico et al. converges spectrally for sufficiently smooth functions i.e. faster than any fixed order in the number of grid points. We implement a performance portable version of the traditional Hockney-Eastwood and the novel Vico-Greengard Poisson solver as part of the IPPL (Independent Parallel Particle Layer) library. For sufficiently smooth source functions, the Vico-Greengard algorithm achieves higher accuracy than the Hockney-Eastwood method with the same grid size, reducing the computational demands of high resolution simulations since one could use coarser grids to achieve them. Additionally, we propose an improvement to the Vico-Greengard method which further reduces its memory footprint. This is important for GPUs, which have limited memory, and should be taken into account when selecting numerical algorithms for performance portable codes. Finally, we showcase performance through GPU and CPU scaling studies on the Perlmutter (NERSC) supercomputer, with efficiencies staying above $50\%$ in the strong scaling case. To showcase portability, we also run the scaling studies on the Alps supercomputer at CSCS, Switzerland and the GPU partition of the Lumi supercomputer at CSC, Finland.

\end{abstract}

\begin{keyword}
Fast free-space Poisson solvers \sep Exascale \sep Performance portability \sep PIC
\end{keyword}

\end{frontmatter}


\clearpage
\section{Introduction}

Simulation codes in many areas of physics need to solve elliptic partial differential equations (PDEs), such as the Poisson equation or the Helmholtz equation. Many computational methods have been developed for this purpose. In particular, we look at integral-based approaches for open boundary conditions, as is customary in previous work \cite{hockney_computer_1988, adelmann_opal_2019, james_solution_1977}. This approach requires dealing with free-space Green's functions, which are singular at the origin and long-range in nature, requiring special attention during computational implementations. 

We take the example of the Poisson equation in the context of electrostatics. The PDE governing our system in this case is given by:
\begin{equation} \label{poisson_eq}
    \Delta \phi(\vec{x}) = - \frac{\rho(\vec{x})}{\epsilon_0}, \\
\end{equation}
where $\rho(\vec{x})$ is the charge density and $\phi(\vec{x})$ the scalar potential, $\epsilon_0$ the vacuum permittivity, and $\vec{x}\in\mathbb{R}^3$. For free-space boundaries ($\phi(\vec{x}) \rightarrow 0 \text{ as } |\vec{x}|\rightarrow \infty$), the solution $\phi$ can be written as a convolution:
\begin{equation} \label{convolution_eq}
    \phi(\vec{x}) = \frac{1}{\epsilon_0}\int G(\vec{x}-\vec{x}')\rho(\vec{x}')d\vec{x}',
\end{equation}
where the Green's function $G(\vec{x})=1/(4\pi|\vec{x}|)$ is the solution of $\Delta \phi = -\delta(\vec{x})$. A computational brute-force approach would consist of discretizing $G$ and $\rho$ on the computational mesh, and approximating the integral as a sum. However, the computational cost of such an operation is $\mathcal{O}(N_g^2)$ \cite{qiang_three-dimensional_2006}, where $N_g=N_xN_yN_z$, with $N_x$, $N_y$, and $N_z$ being the number of grid points in each direction of the 3D grid. For a uniform grid, we have $N_x=N_y=N_z\equiv N$.

For uniform meshes, the computational cost can be reduced by making use of fast Fourier transforms (FFTs).  The convolution theorem states that in Fourier space a convolution can be written as a multiplication \cite{weisstein_convolution_nodate}. \cref{convolution_eq} can therefore be rewritten as $\phi = \mathcal{F}^{-1}\{\mathcal{F}\{G\}\mathcal{F}\{\rho\}\}$ where $\mathcal{F}$ denotes a Fourier transform and $\mathcal{F}^{-1}$ its inverse. This approach drastically reduces the cost to $\mathcal{O}(N_g\text{log}N_g)$. However, FFTs work for periodic signals, and in this case $\rho$ and $G$ are not periodic, since we are interested in free space boundary conditions. Hockney and Eastwood introduced an algorithm which converts the convolution to a cyclic one, making all involved signals periodic, while still obtaining the correct free-space solution at the end \cite{hockney_computer_1988}.

Vico, Greengard, and Ferrando have proposed a novel, fast and spectrally accurate\footnote{Spectrally accurate here refers to the order of convergence.} method which can be applied to many scientific computing problems requiring a convolution with a free-space Green's function \cite{vico_fast_2016}. For smooth distributions, this means as one increases the resolution of the simulation (i.e. the gridsize of the mesh in our computational domain), the solver converges to the exact solution faster than any fixed polynomial order of the gridsize. This would be an improvement in accuracy over the Hockney-Eastwood algorithm, which is only second-order. As suggested in \cite{zou_fft-based_2021}, this method could replace the Hockney-Eastwood algorithm for beam and plasma physics problems, and provide better accuracy with lower resolution and therefore a smaller memory footprint. Using a faster, more accurate, and less memory intensive code is suitable for memory-bound architectures such as GPUs, which are becoming increasingly common in scientific computing and super-computing clusters.

However, the new Vico-Greengard method adds another memory bottleneck, even if there is a gain in accuracy: while in the standard Hockney-Eastwood method one needs to double the domain $(2N)^3$, in the new method a pre-computation of the Green's function on a $(4N)^3$ grid is required to capture the full frequency content \cite{vico_fast_2016}. Storing this $(4N)^3$ grid causes the memory footprint of the Poisson solver to double for each dimension, resulting in an eight-fold memory increase in total, even if it is only done once for a pre-computation. There have been efforts to reduce the fourfold zero-padding factor in literature, for example in \cite{liu_optimal_2024}, the authors show that the optimal zero padding for generic non-local potentials is $\sqrt{d}+1$, where $d$ is the dimension. This reduces the fourfold padding to threefold in the 3D case.

In this work, we further propose an algorithmic improvement by taking advantage of the fact that the truncated spectral Green's function for the specific case of 2D and 3D free space Poisson operator is a real function. Thus by changing from a discrete Fourier transform to the discrete cosine transform, and utilizing the symmetry, we can constrain the pre-computation to a $(2N+1)^3$ grid, reducing the memory and computational costs to be similar to that of the Hockney-Eastwood algorithm. With this improvement, the novel Vico-Greengard method becomes attractive both from the memory requirement as well as its ability to deliver high accuracy solutions. Additionally, this is better for physics cases such as the one presented in \cref{penning_appendix}, where the particle bunch changes during the course of the simulation. In these simulations it is typical to change the computational grid to adapt to the extent of the bunch (known as bunch-fitted domain in the particle accelerator physics community). As the computational grid changes, the Green's function needs to be computed again, as the pre-computed one will not match the new computational domain. As one needs to do this for many time steps ($\mathcal{O}(10^5)$ or more \cite{neveu_parallel_2019}) during the course of the simulation, reducing to a $(2N+1)^3$ grid from a $(4N)^3$ grid would result in much more computational benefits than the static domain cases where it is a one-time cost.

The main contribution of this work is a massively parallel, performance portable, and open-source implementation of the algorithm originally proposed by Vico, Greengard, and Ferrando \cite{vico_fast_2016}, along with modifications to decrease the memory and computational costs. We outline our implementation in Section 2. A comprehensive evaluation of our implementation through convergence studies, scaling studies, and memory benchmarks, demonstrating its superior or equal performance compared to the state-of-the-art Hockney-Eastwood algorithm, is given in Section 3. This implementation of the solvers in the Independent Parallel Particle Layer (IPPL) code also provides users of Particle-Mesh codes with an efficient, massively parallel solver for the free-space Poisson equation which would be suitable for exascale level simulations. For the biggest problem size benchmarked ($1024^3$), the strong scaling studies show an efficiency that stays above $75\%$ both for the GPU and CPU studies on NERSC's Perlmutter supercomputer, reaching runs of about 70\% of the system's capacity (around 4000 GPUs). 

\section{Methodology}
\subsection{IPPL} \label{theory-ippl}

IPPL \cite{frey_ippl-frameworkippl_2023} is an open-source \texttt{C++} framework that provides the tools to develop particle-mesh methods, such as Particle-In-Cell (PIC), explained in \cref{theory-pic}. The framework is built upon Kokkos \cite{carter_edwards_kokkos_2014} which enables cross-platform performance portability for shared memory parallelism. Inter-process communication takes place via the Message Passing Interface (MPI), the de facto standard for distributed memory parallelism. IPPL further depends on
the heFFTe (highly efficient FFT for exascale) library to carry out fast Fourier transforms (FFTs) \cite{ayala_heffte_2020}. Basing IPPL on these libraries makes it a tool for efficient large-scale simulations which can target exascale machines.

IPPL provides application-level users with a modular interface which automatically gives portability and performance when writing simulations within the framework. Its modularity is further enhanced by its dimension independence (can run from 1D to 6D), as well as support for mixed precision.\ More specifically, it is a useful tool for plasma physics, where the particle-in-cell method is commonly used. IPPL has a set of benchmark problems in plasma physics applications, namely ALPINE \cite{muralikrishnan_scaling_2024}, which helps in developing novel algorithms, optimizing HPC implementations, and benchmarking individual components. Further details about the structure of IPPL can be found in \cite{frey_ippl-frameworkippl_2023,muralikrishnan_scaling_2024}.

It is in the context of these particle-mesh methods in IPPL that we seek to solve the Poisson equation (\cref{poisson_eq}). As mentioned previously, for free-space boundary conditions, this is usually done by writing the solution in terms of a convolution, which is efficiently computed in Fourier space and then Fourier transformed back to real space. Until recently, the standard method for doing so was the Hockney-Eastwood method, explained in the following section.

\subsection{Hockney-Eastwood Method}\label{sec:HockneyEastwood}

In order to make the calculation of the convolution in \cref{convolution_eq} efficient, the Hockney-Eastwood method makes the convolution cyclic, such that FFTs can be used. Assuming a 3-dimensional physical domain of size $[0,L_x]\times[0,L_y]\times[0,L_z]$ and a mesh of $N_x\times N_y\times N_z$ grid-points, such that the mesh spacing is $h_x=L_x/N_x$, $h_y=L_y/N_y$, $h_z=L_z/N_z$ in the corresponding directions, the steps of the algorithm are as follows \cite{hockney_computer_1988}:

\begin{enumerate}
    \item Double the computational grid in the non-periodic directions (those which have open boundary conditions). If all boundary conditions are open, this means we will have a grid of size $2N_x \times 2N_y \times 2N_z$.
    
    \item The source term $\rho(\vec{x})$, which is defined on $\vec{x}\in [0,  L_x) \times[0, L_y) \times [0, L_z)$, is then zero-padded to be defined on the full doubled domain. So for $\vec{x} \in [0,  2L_x) \times[0, 2L_y) \times [0, 2L_z)$, we define $\rho_2(\vec{x})$ such that:
    \begin{equation*}
        \rho_2(\vec{x})=\begin{cases}
            \rho(\vec{x}), & \text{if $\vec{x}\in [0,  L_x) \times[0, L_y) \times [0, L_z)$} \\
            0, & \text{otherwise}.
        \end{cases}
    \end{equation*}
    
    \item The Green's function $G(\vec{x})$ also needs to be extended to the doubled domain. This is accomplished by circular shifting and making $G(\vec{x})$ periodic, such that the new function $G_2(\vec{x})$ satisfies:
    \begin{equation*}
        G_2(\vec{x})=\begin{cases}
            G(\vec{x}), & \text{if $\vec{x}\in [0,  L_x) \times[0, L_y) \times [0, L_z)$} \\
            G(2L_x - x, y, z), & \text{if $\vec{x}\in [L_x,  2L_x) \times[0, L_y) \times [0, L_z)$} \\
            G(x, 2L_y - y, z), & \text{if $\vec{x}\in [0,  L_x) \times[L_y, 2L_y) \times [0, L_z)$} \\
            G(x, y, 2L_z - z), & \text{if $\vec{x}\in [0, L_x) \times[0, L_y) \times [L_z, 2L_z)$} \\
            G(2L_x - x, 2L_y - y, z), & \text{if $\vec{x}\in [L_x,  2L_x) \times[L_y, 2L_y) \times [0, L_z)$} \\
            G(2L_x - x, y, 2L_z - z), & \text{if $\vec{x}\in [L_x,  2L_x) \times[0, L_y) \times [L_z, 2L_z)$} \\
            G(x, 2L_y - y, 2L_z - z), & \text{if $\vec{x}\in [0, L_x) \times[L_y, 2L_y) \times [L_z, 2L_z)$} \\
            G(2L_x - x, 2L_y - y, 2L_z - z), & \text{otherwise}.
        \end{cases}
    \end{equation*}
    
    The Green's function is singular at the origin and must therefore be regularized. It is customary to replace the singularity by $G(\vec{0}) = - 1/4\pi$.
    
    \item Since $\rho_2$ and $G_2$ are now periodic with period $(2L_x,2L_y,2L_z)$, we can Fourier transform them (after discretization) using FFTs. In the Fourier domain, the convolution becomes a simple multiplication of the Fourier transforms of $\rho_2$ and $G_2$. In this case, the potential on the doubled domain is given by $\phi_2$:
    \begin{equation*}
       \phi_2 = h_x h_y h_z \mathcal{F}^{-1}\{\mathcal{F}\{\rho_2\} \times \mathcal{F}\{G_2\}\}.
    \end{equation*}
    
    \item The physical solution for the potential is obtained by restricting the double-grid solution to the physical domain, i.e. $\phi(\vec{x}) = \phi_2(\vec{x})$ for $\vec{x} \in [0,  L_x) \times[0, L_y) \times [0, L_z)$. The proof of the equivalence in the physical domain between the cyclic convolution and the original convolution can be found in \cite[p.~12]{ryne_fft-based_2011}.
    
\end{enumerate}

The computational cost to solve the Poisson equation with this method is $\mathcal{O}((8N_g)\text{log}(8N_g))$, where $N_g=N_xN_yN_z$. The trade-off is the increased memory requirement to store the fields $\rho_2$ and $G_2$ on the doubled domain. The Hockney-Eastwood method is second-order accurate, due to the regularization done at $G(\vec{0})$ which makes the Green's function only $C^0$ continuous \cite{rasmussen_particle_2011}. Thus, if high accuracy is desired, the computation quickly becomes expensive. It is therefore advantageous to explore more accurate methods, such as the spectrally accurate Vico-Greengard method \cite{zou_fft-based_2021}, which is briefly summarised below.

\subsection{Vico-Greengard Method}

Vico, Greengard, and Ferrando describe a new method which differs from \cref{sec:HockneyEastwood} by the choice of Green's function. However, this change enables spectral accuracy for smooth functions \cite{vico_fast_2016}. For the Poisson equation, where the differential operator is the Laplace operator, the choice of the Green's function in the Fourier space as per \cite{vico_fast_2016} is
\begin{equation} \label{green_vico}
    G_L(\vec{k}) = 2\left( \frac{\sin(L\abs{\vec{k}}/2)}{\abs{\vec{k}}} \right)^2,
\end{equation}
where $\vec{k}\in\mathbb{R}^3$ is in frequency domain and $L$ is the truncation window size, which must be chosen larger than the maximum distance between any two points in the computational domain \cite{vico_fast_2016}, e.g. $L = \alpha \sqrt{L_x^2 + L_y^2 + L_z^2}$ with $\alpha > 1$ for a domain $[0, L_x]\times[0, L_y]\times[0,L_z]$. Here, we have chosen $\alpha=1.1$.

To avoid any loss of information when calculating the inverse Fourier transform of \cref{green_vico}, the transform must be performed on a $4N_x \times 4N_y \times 4N_z$ grid due to the oscillatory nature of $G_L$. First, as explained by Hockney and Eastwood \cite[p.~213]{hockney_computer_1988}), we need to have a grid of double the original size to perform the aperiodic convolution.
The additional factor of two in the mesh size compared to the Hockney-Eastwood method comes from the fact that $G_L$ is an oscillatory signal. As stated by the Nyquist sampling theorem, an oscillatory signal must be sampled at a sampling rate $f_s$ two times larger than the maximum frequency $f_{\max}$ of that signal, i.e. $f_s > 2f_{\max}$, to not lose any information.

However, if all the convolution computation is performed on the quadruple grid, the source term must also be zero-padded on the quadruple domain, which makes the method prohibitively memory-intensive for computational simulations. Fortunately, this can be avoided by calculating the inverse transform of $G_L$ on the fourfold grid in only one pre-computation step, which is then restricted to the doubled grid of size $2N_x \times 2N_y \times 2N_z$ to obtain $G_2$. The restricted Green's function $G_2$ is then reused every time step in which the Poisson solve is called during the simulation. The quadruple grid is therefore never needed again during the course of the simulation unless the grid spacing changes, in which case the Green's function needs to be updated. Thus, the algorithm follows the same steps as in \cref{sec:HockneyEastwood} except for the Green's function computation, where we use the pre-computed $G_2$. The complexity is similar to the Hockney-Eastwood method, but the method has a higher accuracy for sufficiently smooth functions. The main drawback is the need for a $4N_x \times 4N_y \times 4N_z$ grid in the pre-computation step, which increases memory demands by a factor of eight compared to the Hockney-Eastwood method. This is especially a problem with GPU architectures, which generally have less memory available than their CPU counterparts. In order to circumvent this, we present an algorithmic change to reduce the memory required by the pre-computation step.

\subsection{A Modified Vico-Greengard Method}

A simple modification in the above algorithm allows us to reduce the grid size of the pre-computation step from $(4N_x \times 4N_y \times 4N_z)$ to $((2N_x +1 ) \times (2N_y + 1) \times (2N_z + 1))$. We make use of the fact that the Green's function $G_L$ used in the Vico-Greengard method, given by \cref{green_vico}, is purely real and has even symmetry i.e. $G_L(\vec{k}) = G_L(-\vec{k})$. In view of the even symmetry, the Green's function on the Fourier domain $G_L$, which is defined on a $4N_x \times 4N_y \times 4N_z$ grid to contain the full frequency content, actually only has $((2N_x + 1) \times (2N_y + 1) \times (2N_z + 1))$ unique values. Furthermore, the discrete Fourier transform reduces to a discrete cosine transform (DCT) thanks to the same properties of the signal, namely that $G_L$ is purely real and even. In this way, we can perform the pre-computation step of the Vico-Greengard method on the $(2N+1)^3$ domain instead of the fourfold domain, and use the DCT to compute the inverse transform of $G_L$. This way, the memory footprint of the Vico-Greengard method is the same order-of-magnitude  to the Hockney-Eastwood method. Implementation-wise, we use the discrete cosine transform which is available in heFFTe \cite{montanaro_improvements_2023}.

\subsection{Implementation specifics} \label{sec:implementation}

In the case of the free-space convolution-based Poisson solver as presented above, the computational implementation includes moving the field from the original computational grid of size $N^3$ to a $(2N)^3$ grid, and vice-versa. In the Vico-Greengard method, there is also a restriction step required from the $(4N)^3$ grid to the $(2N)^3$ grid (or $(2N+1)^3$ to $(2N)^3$ in the modified DCT case). This requires a non-trivial implementation in massively parallel codes due to the domain decomposition of the computational grid, which incurs communication costs to exchange field data among processors. These implementation details are explained in \cref{MPI_implementation}. The goal is to have an efficient implementation of the communication routine which does not impact the performance and scalability of the solver, such that it is only bounded by the performance of the FFTs.

\section{Results and Discussion}
\subsection{Correctness test and accuracy analysis}

To test the correctness and accuracy of the numerical methods used in our implementation, we use the analytical solution of the Poisson's equation for a Gaussian source with mean $\mu$ and standard deviation $\sigma$
\begin{equation}\label{gaussian}
    \rho = -\frac{1}{\sqrt{(2\pi)^3}}\exp\left(- \frac{r^2}{2\sigma^2} \right), 
\end{equation}
which is given by \cite{zou_fft-based_2021}
\begin{equation*}
    \phi_{exact} = \frac{1}{4\pi r}\text{erf}\left( \frac{r}{\sqrt{2}\sigma} \right),
\end{equation*}
where $\text{erf}$ is the error function and $r = \sqrt{(x-\mu)^2+(y-\mu)^2+(z-\mu)^2}$. We position the Gaussian in the center of a unit box, i.e. $\mu=0.5$ with $L_x=L_y=L_z=1$, and $\sigma=0.05$.

In a second test case, we show the behaviour of the numerical schemes when the source is non-smooth. The test case described in \cite{budiardja_parallel_2011} has a source term given by
\begin{equation}\label{sphere}
\rho = \begin{cases}
          4\pi G, & \text{if $r \leq$ 1}\\
          0, & \text{if $r > $ 1}
        \end{cases},
\end{equation}
and the exact solution
\begin{equation*}
\phi_{exact} =\begin{cases}
			     -\frac{2}{3}\pi G (3 - r^2), & \text{if $r \leq$ 1}\\
                 -\frac{4}{3} \pi \frac{G}{r}, & \text{if $r > $ 1}
		       \end{cases},
\end{equation*}
where $G=6.67408\cdot 10^{-11}$ m$^3$kg$^{-1}$s$^{-2}$ is the gravitational constant and $r$ is as above. We choose $\mu = 1.2$ to center the distribution in the 3D cubic box of extents $L_x=L_y=L_z=2.4$. The solution $\phi$ is the gravitational potential of a sphere.

\begin{figure}[h]
    \centering
    \includegraphics[width=0.9\textwidth]{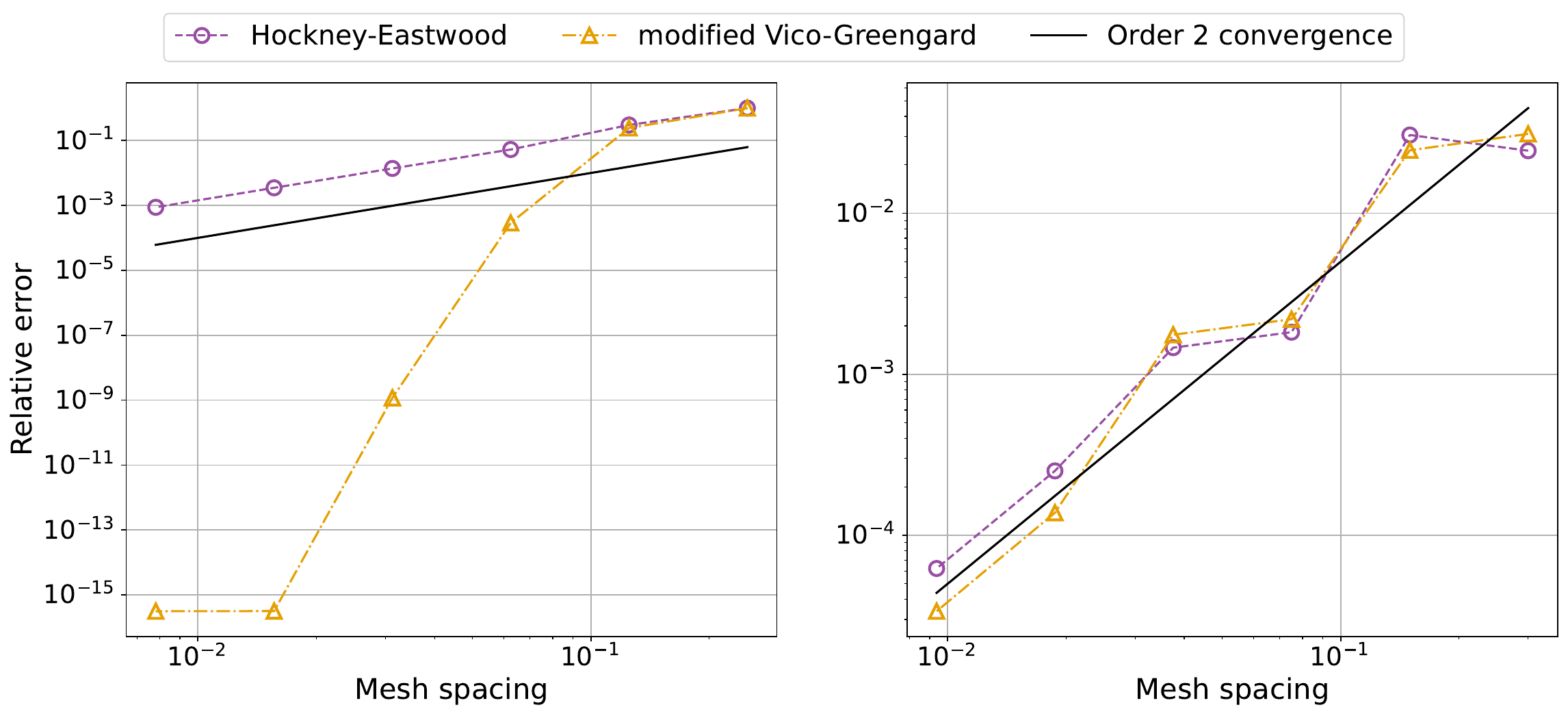}
    \caption{Relative error between analytical and computed solutions for two of the solvers (Hockney-Eastwood and the modified version of Vico-Greengard). The original Vico-Greengard follows the same line as the modified one, but has been omitted from the plot for readability.} The 2nd-order convergence theoretical line is plotted for reference. Left: Source term given by \cref{gaussian}. Right: Source term given by \cref{sphere}.
    \label{convergence_plot}
\end{figure}

In order to compare the numerical and analytical solutions, we use the L2 norm relative error between the computed solution and the exact solution
\begin{equation*}
    \frac{||\phi - \phi_{exact}||_2}{||\phi_{exact}||_2}, 
\end{equation*}
where the L2 norm is taken over the values of $\phi$ on the discrete mesh points in our computational domain.
By means of a convergence study we compare the error convergence rates with their theoretical values to verify the correctness of the implementation. In \cref{convergence_plot}, we obtain the expected second-order convergence with the Hockney-Eastwood solver, whereas with the modified Vico-Greengard solver we get spectral convergence for the smooth source and second-order convergence for the non-smooth source terms, which follows the same trend as the original, proving the correctness of our algorithmic modifications.

These results also shed light on the limitations of the new solver: the convergence of Vico-Greengard solver depends on the smoothness of the source term. For the second test case with a non-smooth source term, i.e. \cref{sphere}, the Vico-Greengard solver is only as good as the Hockney-Eastwood solver. However, since the modified Vico-Greengard solver does not cause any additional computational costs in terms of memory and complexity compared to the Hockney-Eastwood solver, it can replace it as the default method. The accuracy is therefore limited by the properties of the source term instead of the solver itself. 

\subsection{Scaling studies}

We run strong and weak scaling studies of both the Hockney-Eastwood and Vico-Greengard solvers implemented in IPPL. To showcase portability and scaling across architectures, these studies are conducted on both the CPU and GPU partitions of the Perlmutter supercomputer at NERSC (USA), as well as on the Alps supercomputer at CSCS (Switzerland), and the GPU partition of the EuroHPC Lumi supercomputer at CSC (Finland), all of them from HPE Cray.

The GPU partition of Perlmutter contains 1536 nodes where each of them is equipped with 4 Nvidia A100 GPUs. The CPU partition consists of 3072 nodes, with 2 AMD EPYC 7763 CPUs of 64 cores each. We run the simulations with 1 MPI rank per GPU for the GPU scaling studies and 128 MPI ranks per CPU node (1 MPI task per physical core) for the CPU scaling studies. As for Alps, it has 2688 nodes with 4 sockets each with the Nvidia Grace-Hopper (GH200) architecture, and we run the simulations with 1 MPI rank per socket. The last architecture we test on is AMD-based, on the GPU partition of the Lumi supercomputer, which contains 2978 nodes with 4 AMD MI250X GPUs with two compute dies per node, so effectively 8 GPUs per node.

The strong scaling study is done by keeping the problem size fixed and measuring the timing as we increase the number of nodes for the simulation. We run this for two different problem sizes: $512^3$ and $1024^3$. The weak scaling study consists in increasing the number of nodes used for the simulation while maintaining a constant workload, i.e. we increase the problem size with the number of nodes. Starting with $512^3$ as our problem size, we increase to $512^2\times1024$ when doubling the number of nodes, then $512\times1024^2$ for four times the initial node count and finally ending at $1024^3$. The ``Solve" time shown in the studies is the total runtime of the simulation without the time spent on the setup of the solver (the initialization time, which includes the FFT workspace setup and the field allocations). The memory requirements for the problem sizes we are studying are shown in \cref{memory_table}, for both the Hockney-Eastwood and the modified Vico-Greengard solvers.
\setlength{\tabcolsep}{3pt} % Default value: 6pt
\renewcommand{\arraystretch}{1.2} % Default value: 1
\begin{table}[h]
\centering
\begin{tabular}{ |c|c|c| } 
 \hline
 Method & $512^3$ size & $1024^3$ size \\
 \hline
 Hockney-Eastwood & 60 GB & 480 GB \\ 
 Modified Vico-Greengard & 87 GB & 700 GB \\ 
 \hline
\end{tabular}
\caption{Memory footprint of the two solvers of the two relevant problem sizes for the scaling studies, measured on GPU the Alps machine using the Kokkos profiling tools \cite{kokkoskokkos-tools_kokkos_2025}.}
\label{memory_table}
\end{table}

\subsubsection{Perlmutter}

On Perlmutter, IPPL was compiled with Cuda 12.0 (GPU-specific), FFTW 3.3.10.5 (CPU-specific), Kokkos 4.1.0, Heffte 2.4.0, and manual installations of MPICH (due to some technical issues, a manual installation of MPICH was needed in order to be able to compile with Cuda 12 on the cluster). One small caveat is that the MPICH version used to compile IPPL on CPU is 4.1.2 whereas for GPU it is 4.1.1, due to some technical issues on the cluster. Since this is a minor version change with bug fixes related to ``user-reported crashes and build issues" \cite{raffenetti_mpich_nodate}, we do not expect differences in our scaling studies.

\begin{figure}[H]
\centering
\begin{subfigure}{.425\textwidth}
  \centering
  \includegraphics[width=\linewidth]{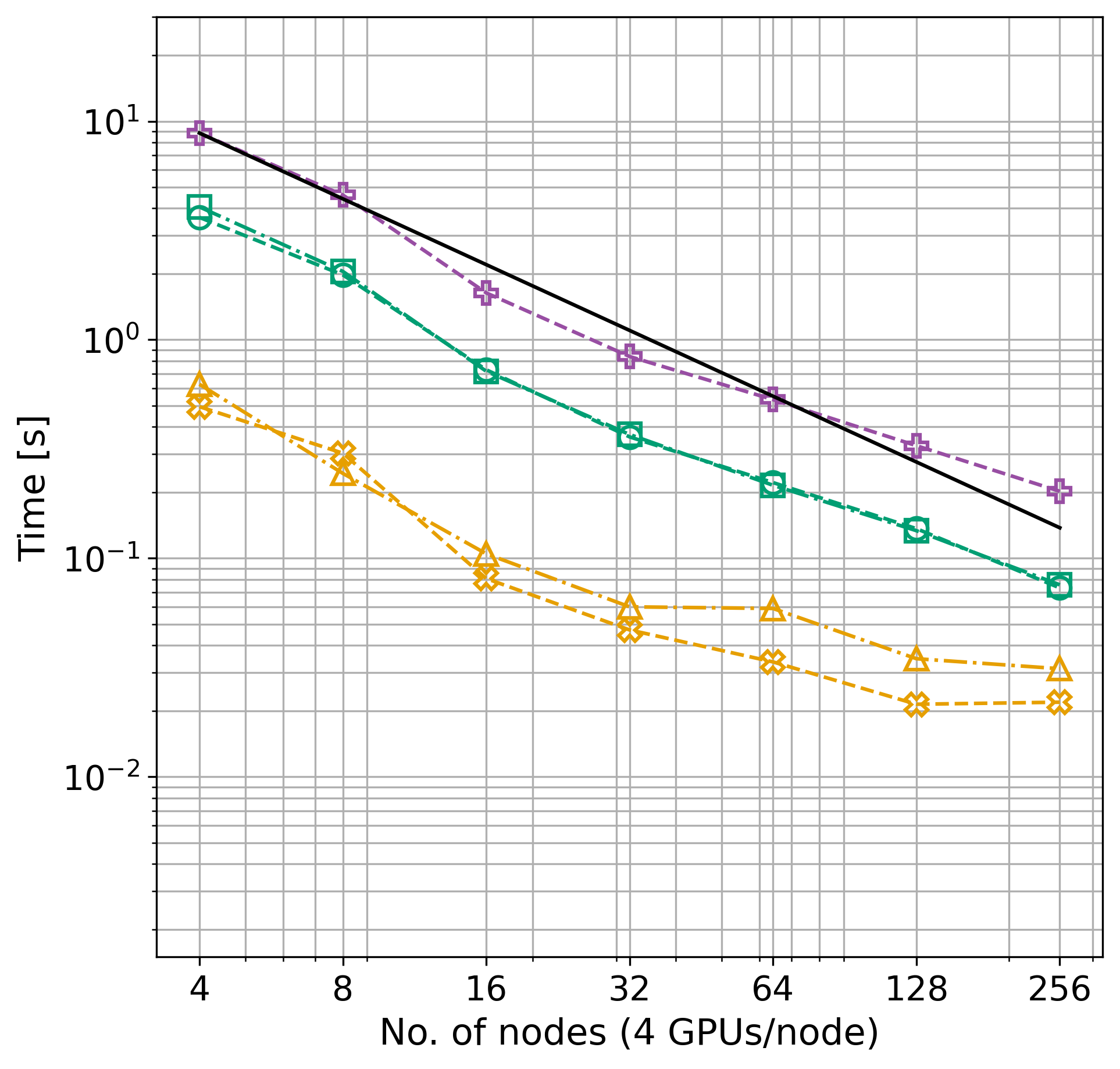}
\end{subfigure}\hfill
\begin{subfigure}{.55\textwidth}
  \centering
  \includegraphics[width=\linewidth]{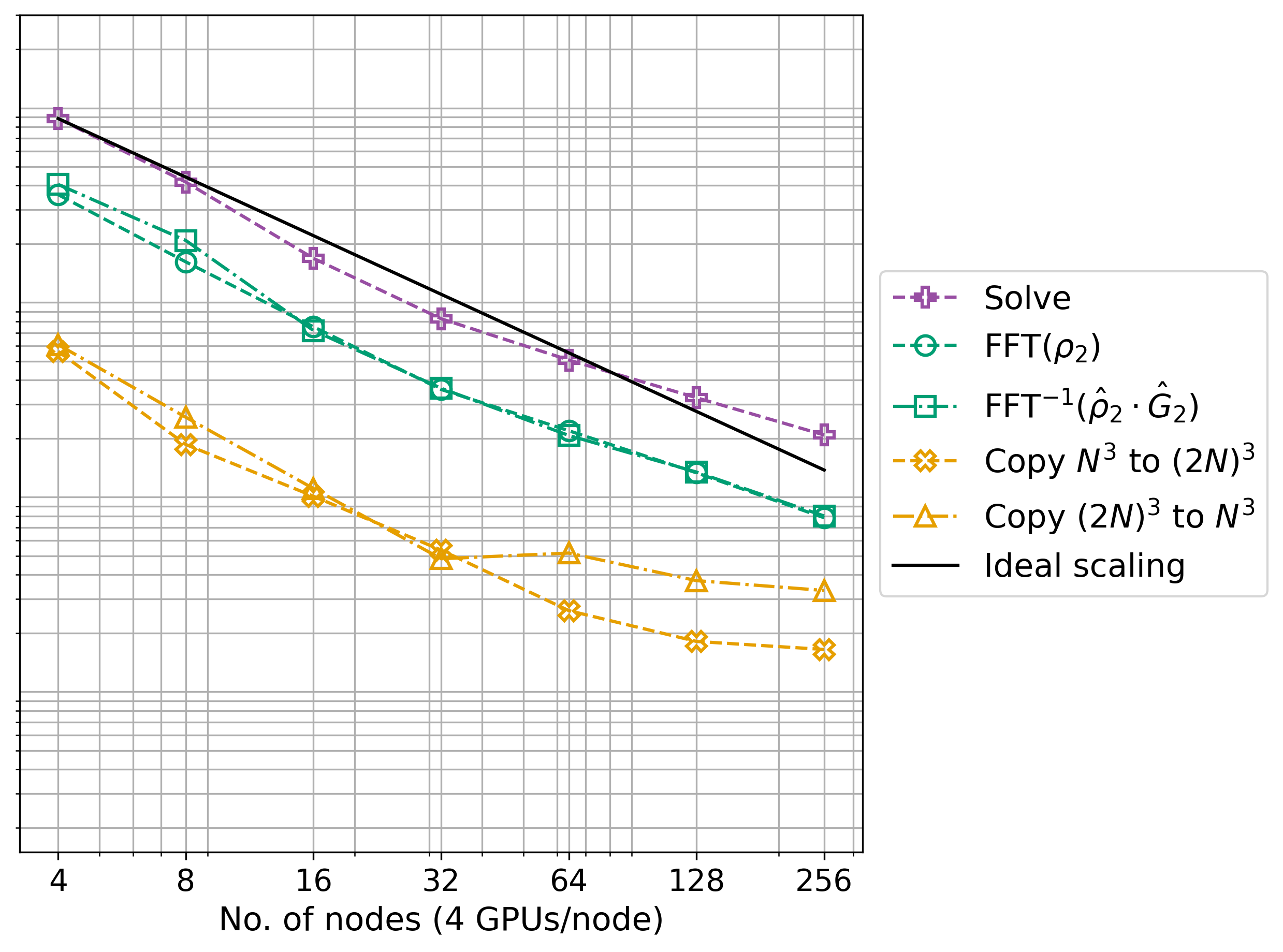}
\end{subfigure}
\caption{Strong scaling results on Perlmutter GPU for both the Hockney-Eastwood (left) and the modified Vico-Greengard solver (right) with a problem size of $N^3 = 512^3$, using the \texttt{heFFTe} parameters of pencil decomposition, a2av communication, no reordering, and with GPU-aware enabled. The efficiency stays above 65\%.}
\label{512_strong_scaling_gpu}
\end{figure}

In \cref{512_strong_scaling_gpu} we show the GPU strong scaling results of the $512^3$ grid simulation for both the Hockney-Eastwood algorithm (left panel) and the modified Vico algorithm (right panel). The purple lines denote the overall algorithm execution times (``Solve"). The timings of the forward and backward FFT transforms are highlighted in green. The costs for the data transfer between the grids of size $N^3$ and $(2N)^3$ and vice versa are shown in yellow. The algorithms have nearly identical execution times and they show very good scaling behaviour up to 256 nodes (i.e.\ 1024 GPUs). The parallel efficiency does not drop below 65\%. After 32 nodes, the data copy kernel timings (red lines) start to flatten since the communication costs associated with the data transfer break even with the actual data copy. We see that the the costs of the copy between grids stay an order of magnitude lower in absolute timings than the FFT routines. The solvers stay FFT dominated, as intended (see \cref{sec:implementation}).

\begin{figure}[H]
\centering
\begin{subfigure}{.425\textwidth}
  \centering
  \includegraphics[width=\linewidth]{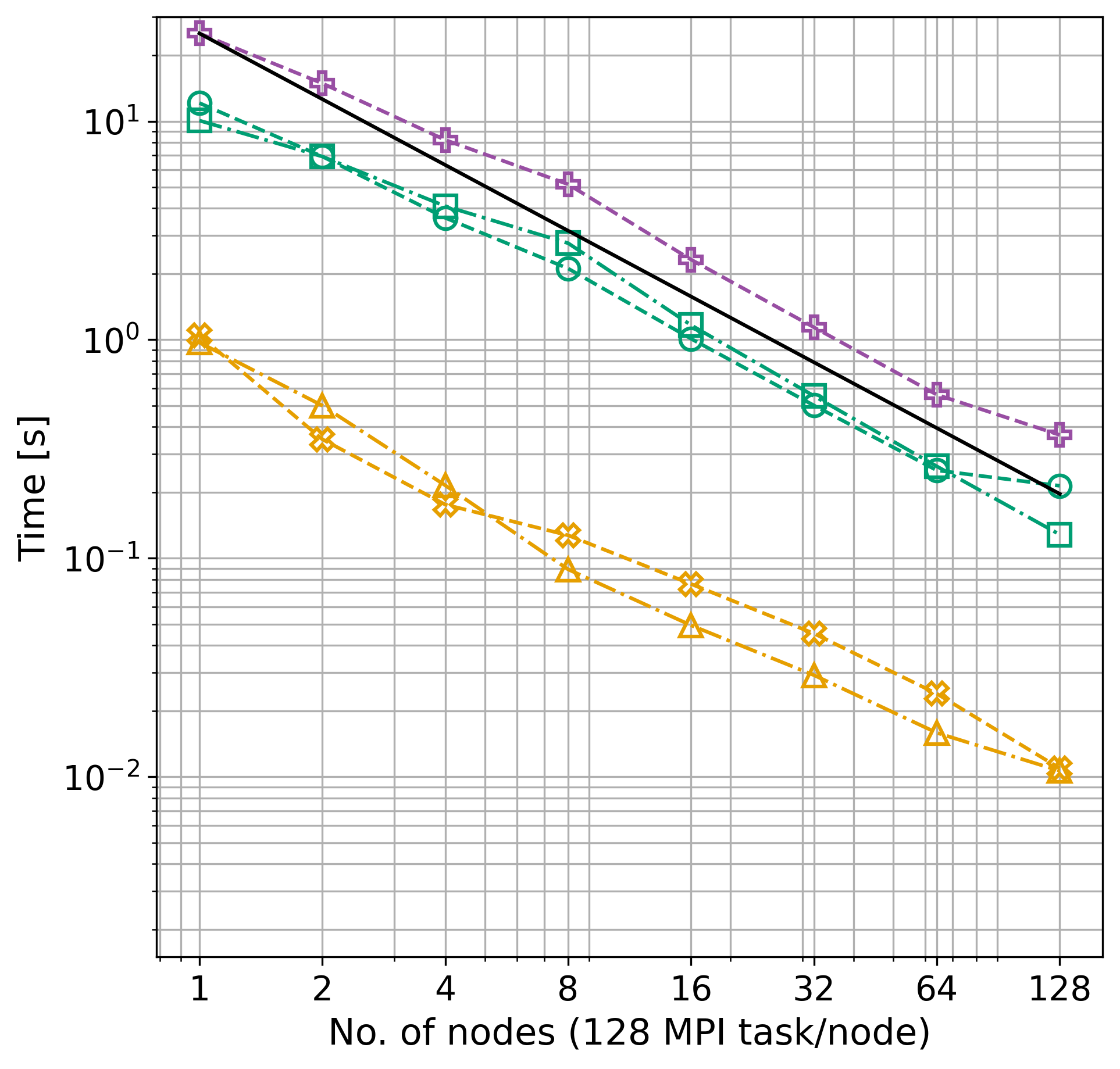}
\end{subfigure}\hfill
\begin{subfigure}{.55\textwidth}
  \centering
  \includegraphics[width=\linewidth]{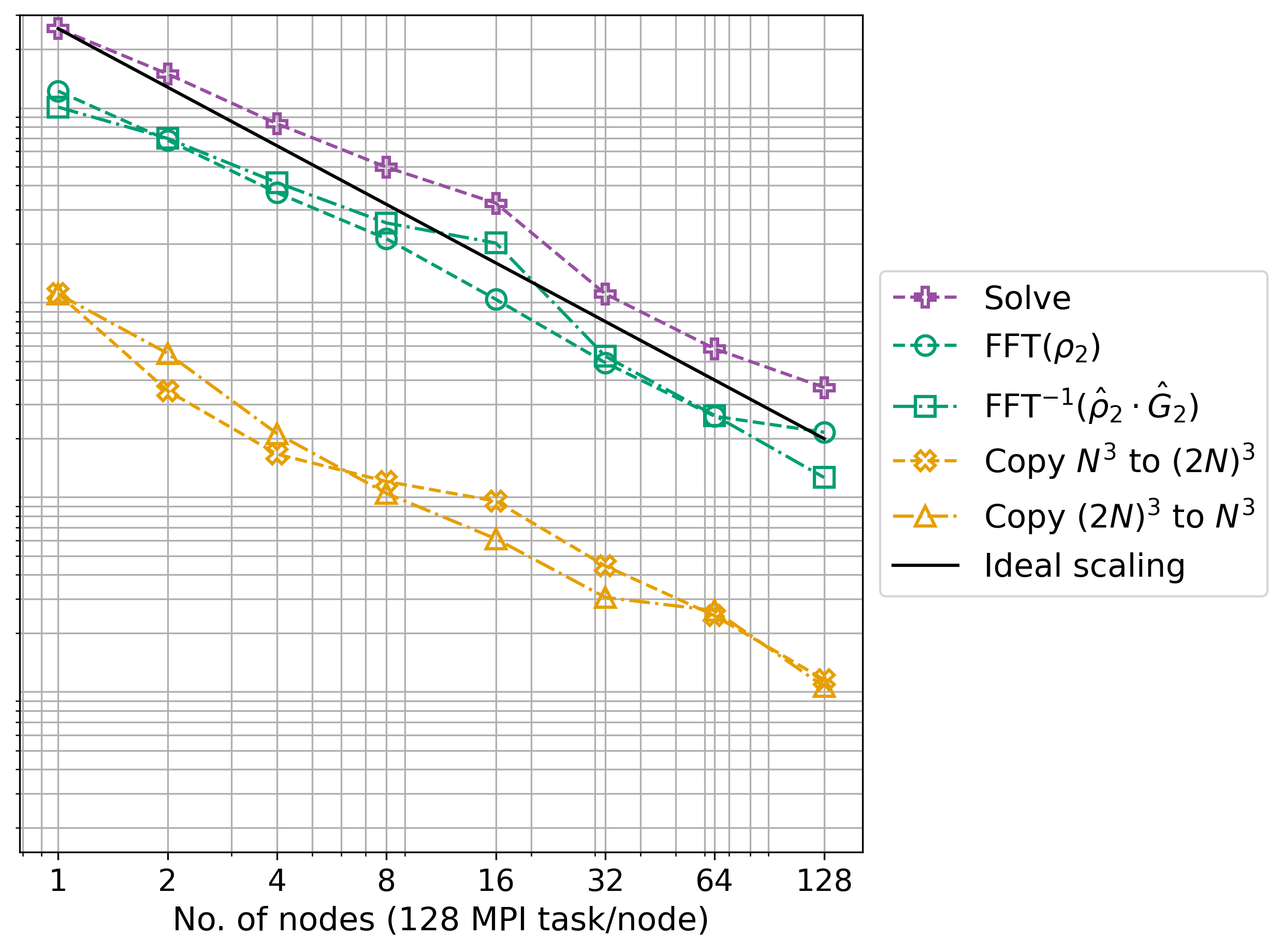}
\end{subfigure}
\caption{Strong scaling results on Perlmutter CPU for both the Hockney-Eastwood (left) and the modified Vico-Greengard solver (right) with a problem size of $N^3 = 512^3$, using the \texttt{heFFTe} parameters of pencil decomposition, a2av communication, no reordering. The efficiency stays above 50\%.}
\label{512_strong_scaling_cpu}
\end{figure}

The CPU strong scaling of the same setup is given in \cref{512_strong_scaling_cpu}. We again observe very good scaling behaviour, i.e.\ the parallel efficiency remains above 50\%. Note that here we start with 1 node rather than 4 nodes as for the GPU scaling due to the smaller amount of memory on the GPUs. Also, the data transfer timings do not flatten as we have seen for the GPU runs because the actual data copy on the CPU is slower than on the GPU and therefore hides the associated communication costs. The dominating cost on both hardware systems (GPU and CPU) are the FFT transforms, as expected. Note that the FFT timers also include communication costs.


\begin{figure}[H]
\centering
\begin{subfigure}{.425\textwidth}
  \centering
  \includegraphics[width=\linewidth]{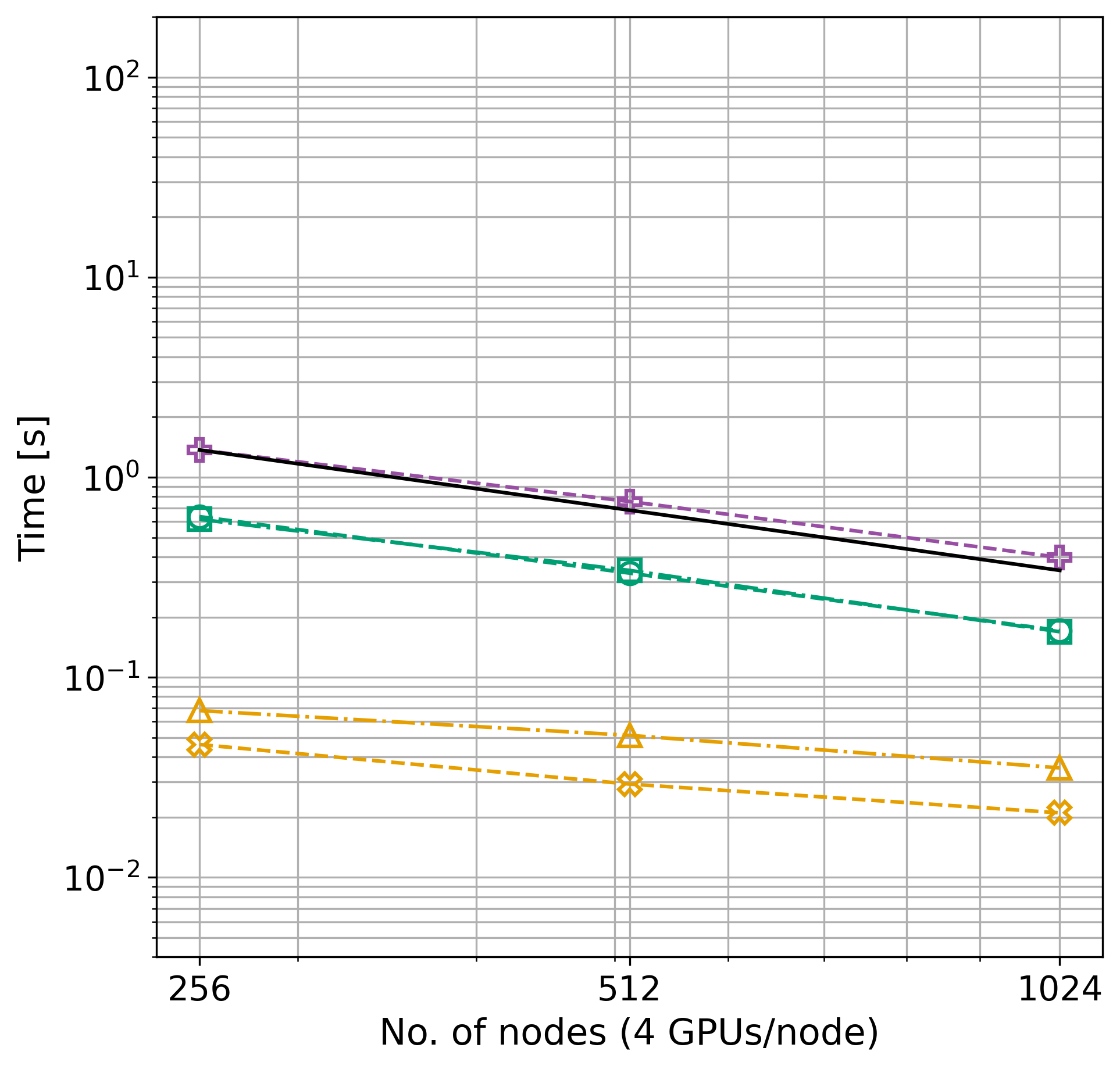}
\end{subfigure}\hfill
\begin{subfigure}{.55\textwidth}
  \centering
  \includegraphics[width=\linewidth]{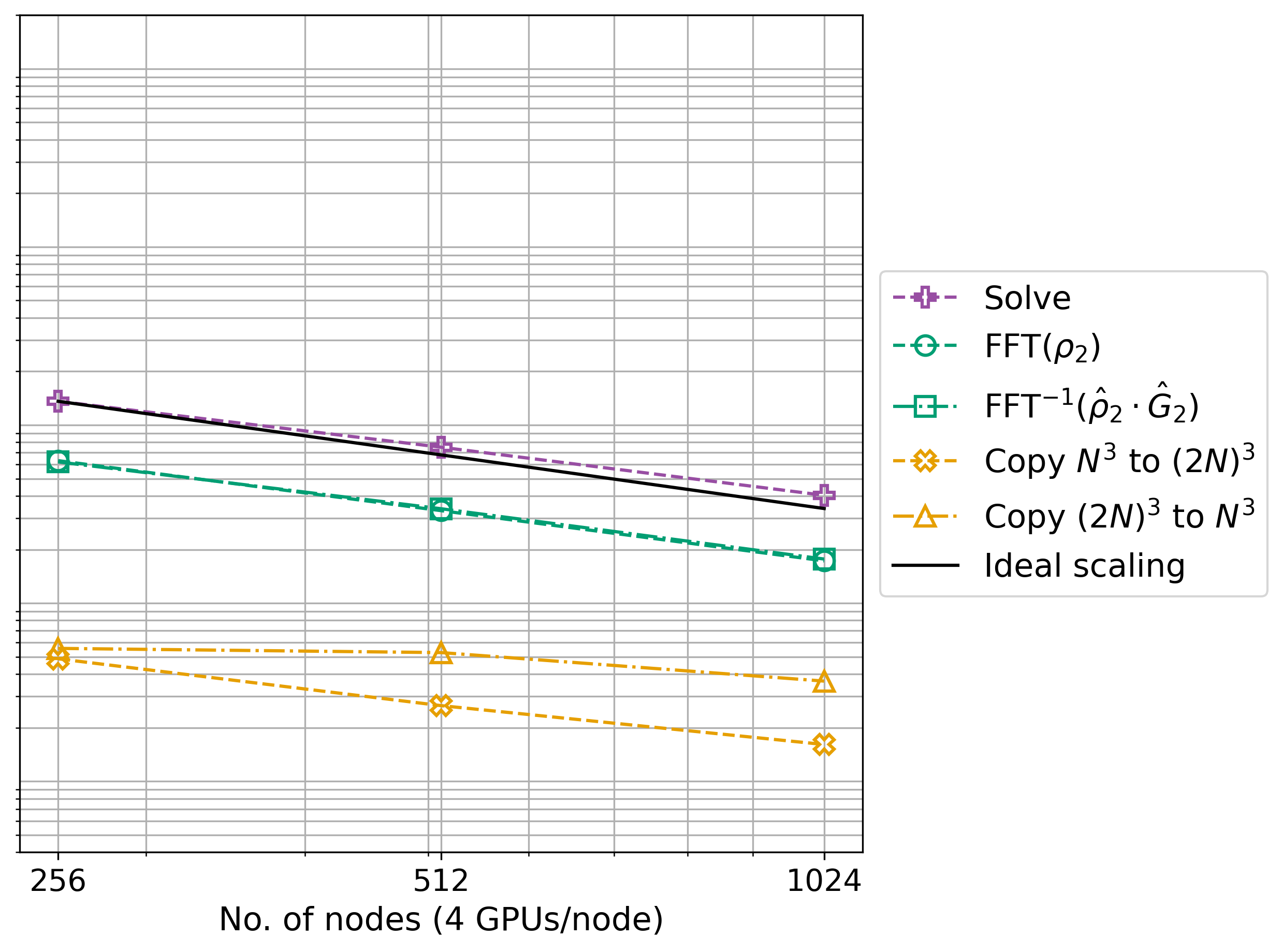}
\end{subfigure}
\caption{Strong scaling results on Perlmutter GPU for both the Hockney (left) and the modified Vico solver (right) with a problem size of $N^3 = 1024^3$, using the \texttt{heFFTe} parameters of pencil decomposition, a2av communication, no reordering, and with GPU-aware enabled. The efficiency stays above 80\%.}
\label{1024_strong_scaling_gpu}
\end{figure}

\begin{figure}[H]
\centering
\begin{subfigure}{.425\textwidth}
  \centering
  \includegraphics[width=\linewidth]{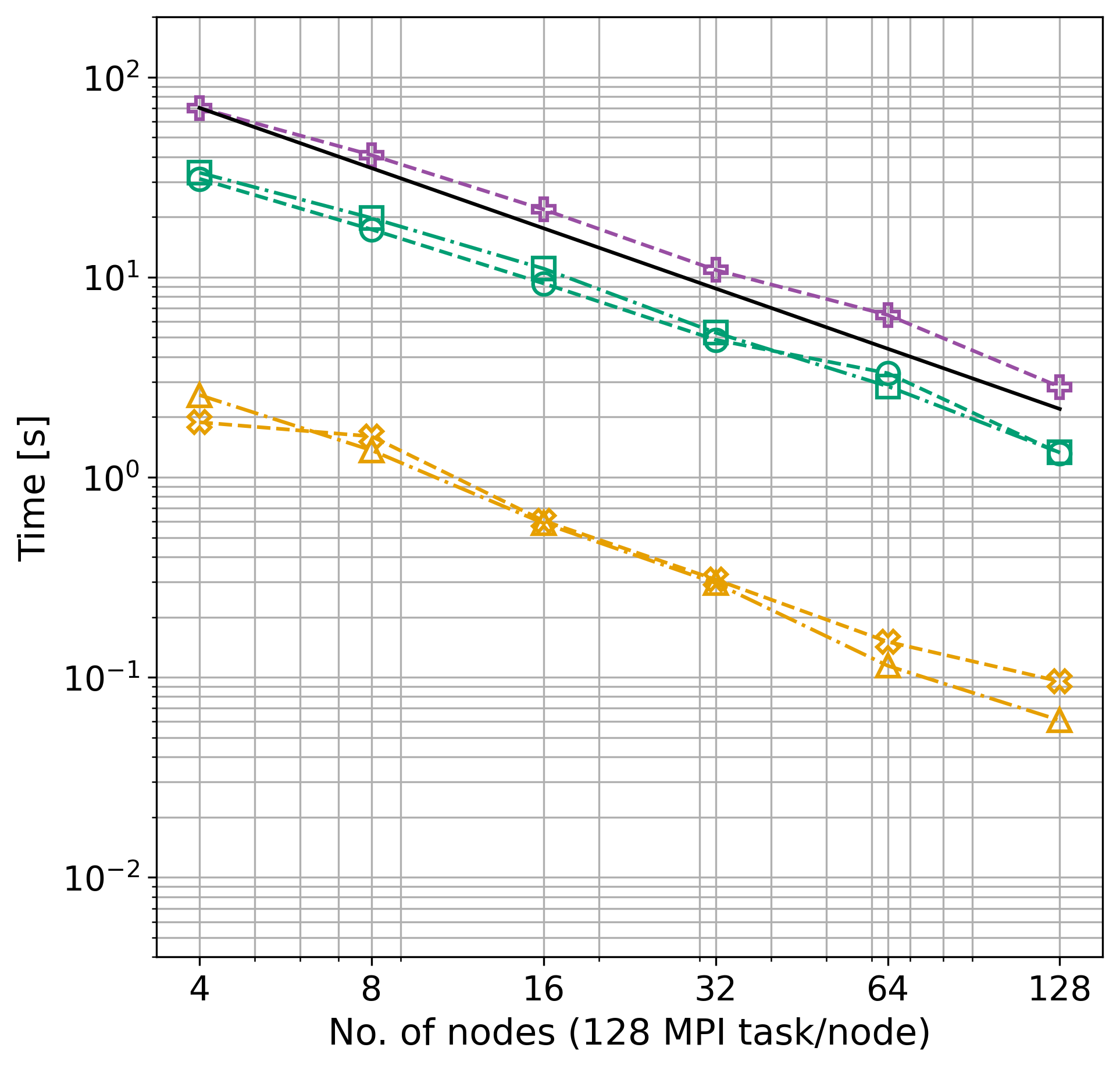}
\end{subfigure}\hfill
\begin{subfigure}{.55\textwidth}
  \centering
  \includegraphics[width=\linewidth]{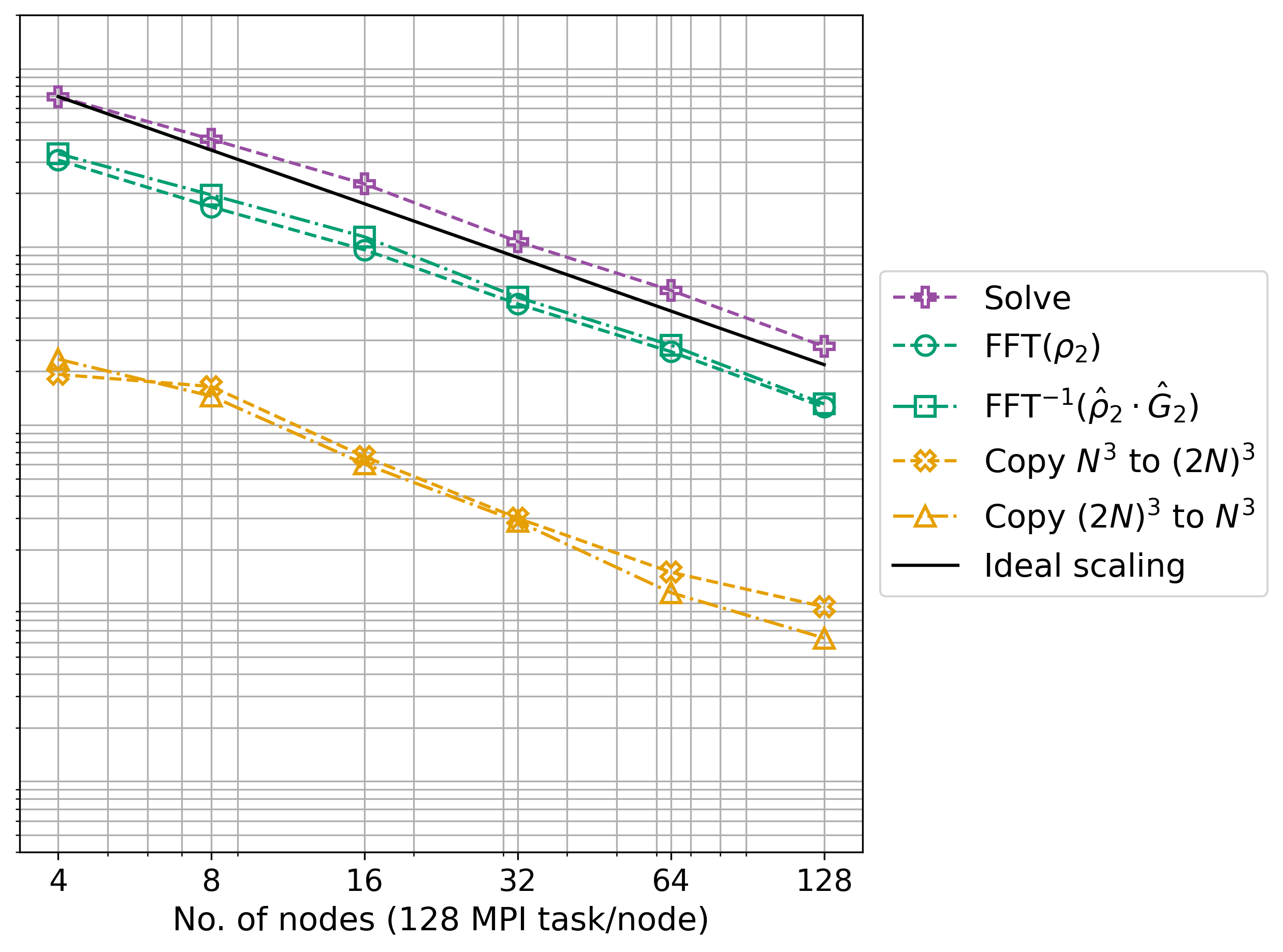}
\end{subfigure}
\caption{Strong scaling results on Perlmutter CPU for both the Hockney-Eastwood (left) and the modified Vico-Greengard solver (right) with a problem size of $N^3 = 1024^3$, using the \texttt{heFFTe} parameters of pencil decomposition, a2av communication, no reordering. The efficiency stays above 65\%.}
\label{1024_strong_scaling_cpu}
\end{figure}

The GPU and CPU strong scalings for a problem size of $1024^3$ grid points are provided in \cref{1024_strong_scaling_gpu} and \cref{1024_strong_scaling_cpu}, respectively. We obtain similar scaling behaviour on both hardware systems as for $512^3$ grid points and the parallel efficiency remains above 65\%.


\begin{figure}[H]
\centering
\begin{subfigure}{.425\textwidth}
  \centering
  \includegraphics[width=\linewidth]{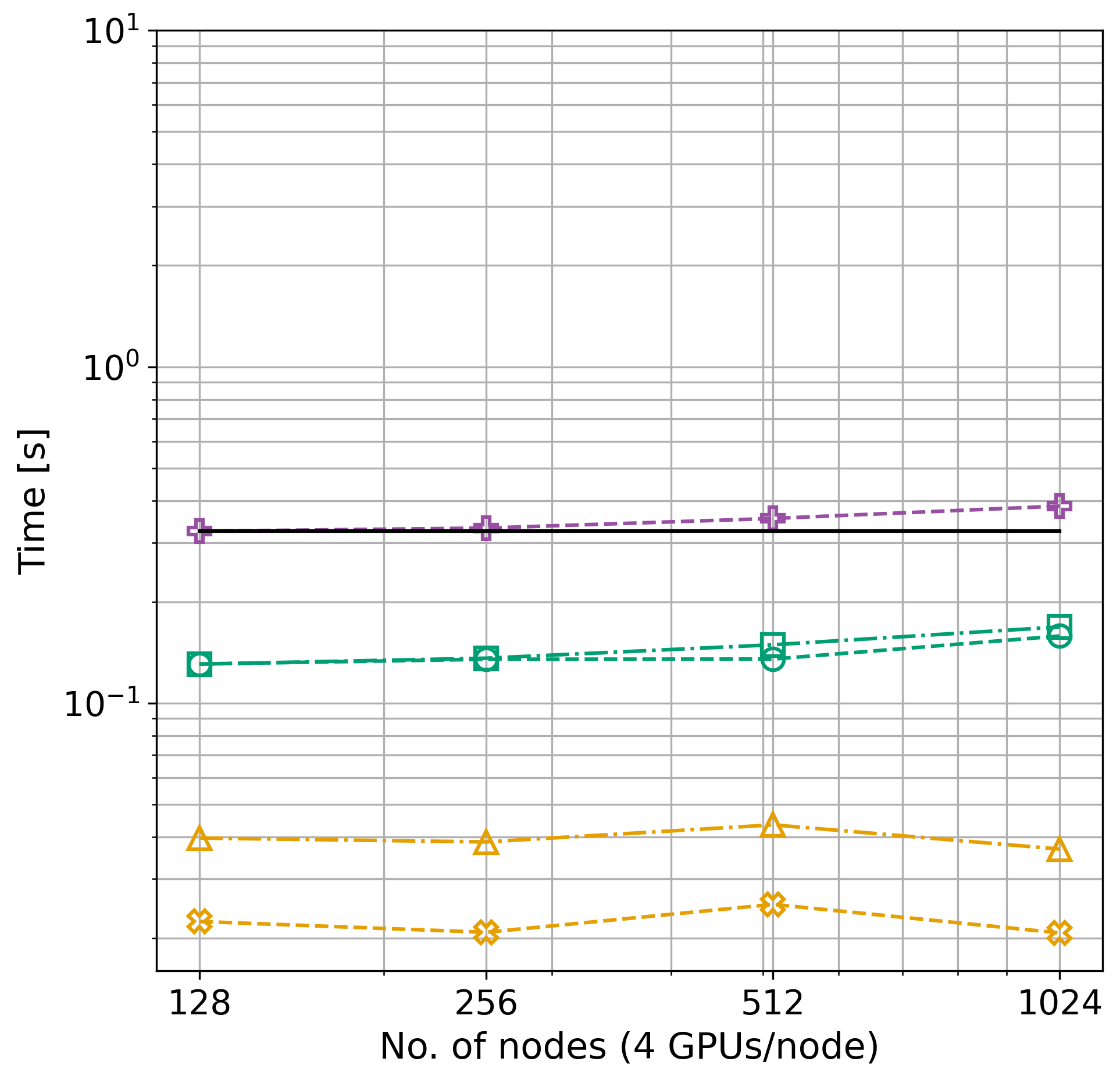}
\end{subfigure}\hfill
\begin{subfigure}{.55\textwidth}
  \centering
  \includegraphics[width=\linewidth]{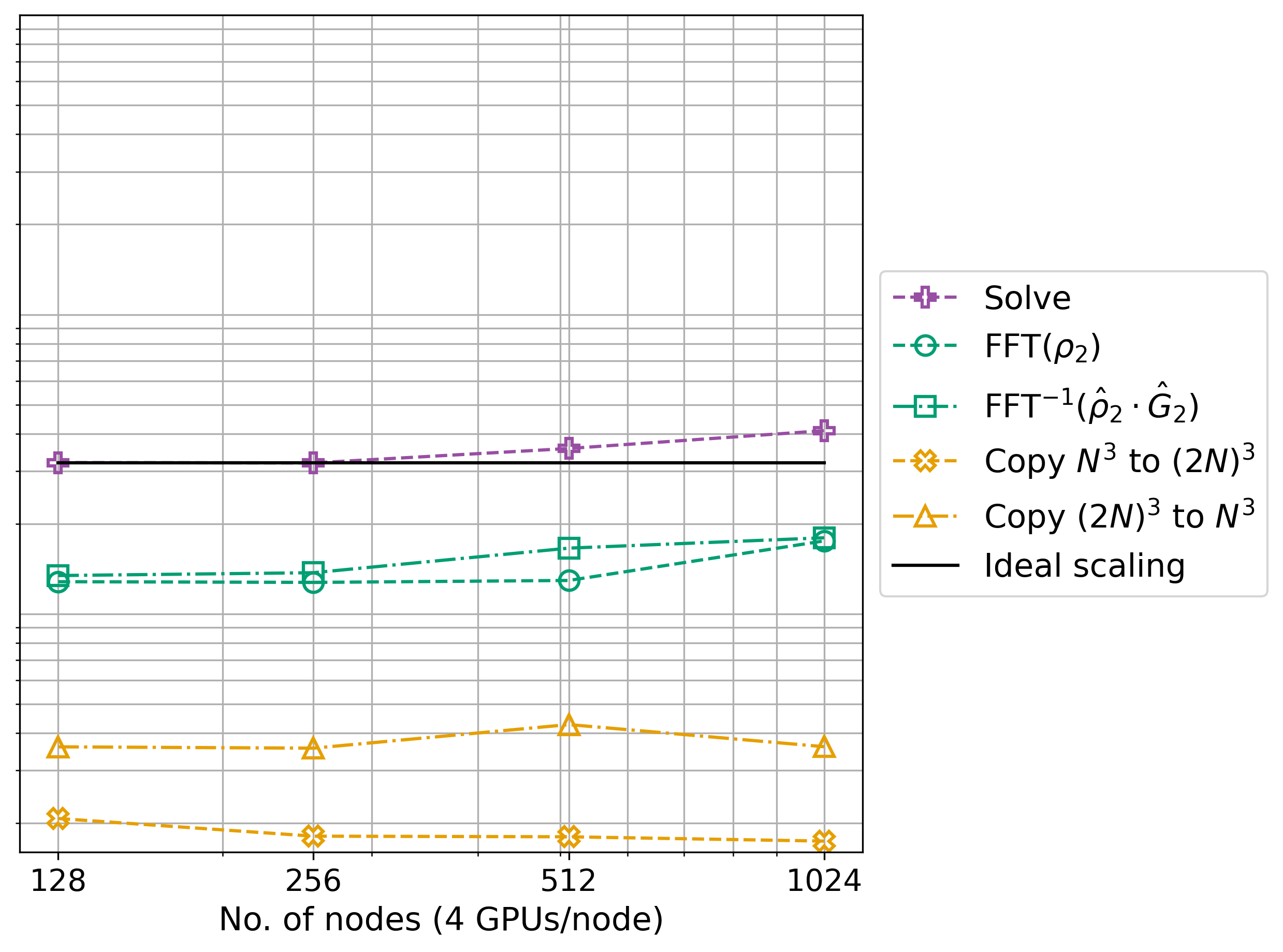}
\end{subfigure}
\caption{Weak scaling results on Perlmutter GPU for both the Hockney-Eastwood (left) and the modified Vico-Greengard solver (right), starting from a problem size of $N^3 = 512^3$ and increasing the workload proportionally to the node increase until reaching a problem size of $N^3 = 1024^3$. The \texttt{heFFTe} parameters used are pencil decomposition, a2av communication, no reordering, and GPU-aware enabled. The efficiency stays above 75\%.}
\label{weak_scaling_gpu}
\end{figure}

\begin{figure}[H]
\centering
\begin{subfigure}{.425\textwidth}
  \centering
  \includegraphics[width=\linewidth]{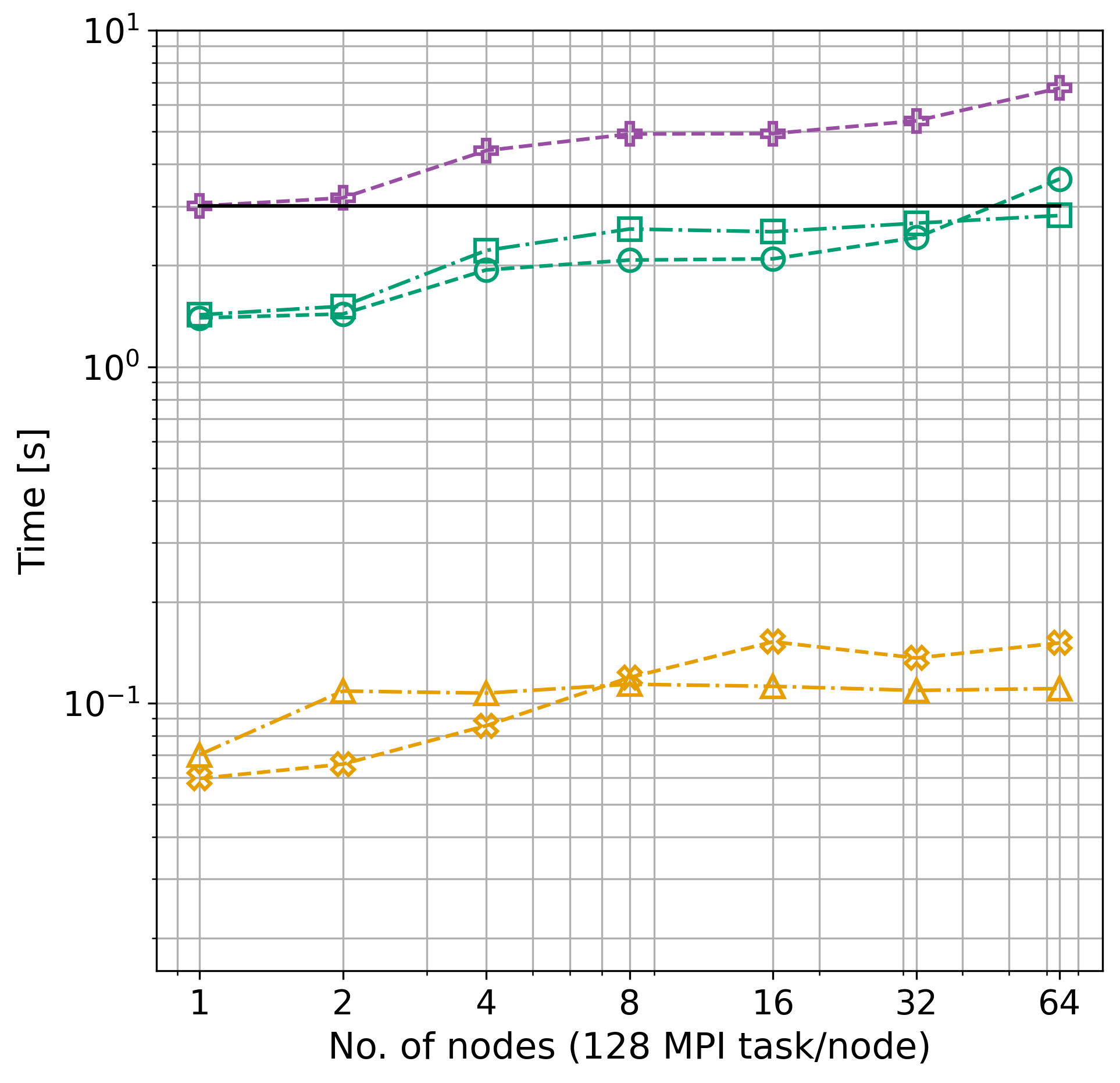}
\end{subfigure}\hfill
\begin{subfigure}{.55\textwidth}
  \centering
  \includegraphics[width=\linewidth]{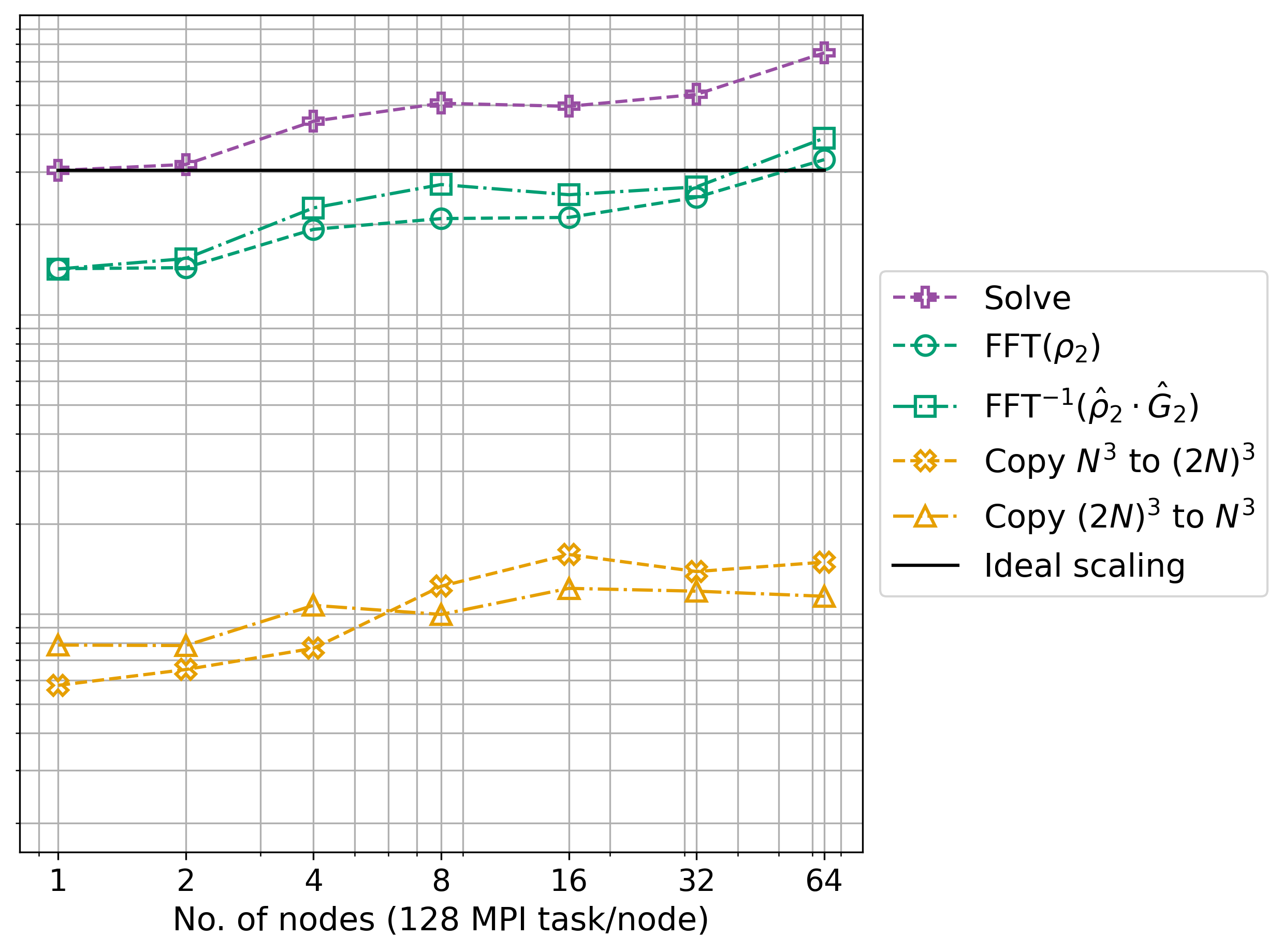}
\end{subfigure}
\caption{Weak scaling results on Perlmutter CPU for both the Hockney-Eastwood (left) and the modified Vico-Greengard solver (right), starting from a problem size of $N^3 = 512^3$ and increasing the workload proportionally to the increase in node counts until reaching a problem size of $N^3 = 1024^3$. The \texttt{heFFTe} parameters used are pencil decomposition, a2av communication, no reordering. The efficiency goes down to $40\%$.}
\label{weak_scaling_cpu}
\end{figure}

The weak scaling results on both hardware systems are given in \cref{weak_scaling_gpu} and \cref{weak_scaling_cpu}. We start at a problem size of $512^3$ grid points and increase the problem size up to $1024^3$ grid point while keeping the workload per node approximately constant. On both hardware systems we experience a loss in parallel efficiency because the cost of the all-to-all communication in the FFTs increases with the node count while the workload is kept constant. For more details on the scaling of the FFTs, we refer the reader to studies conducted by the heFFTe library team \cite{ayala_analysis_2022}.


\subsubsection{Alps}

\begin{figure}[H]
\centering
\begin{subfigure}{.425\textwidth}
  \centering
  \includegraphics[width=\linewidth]{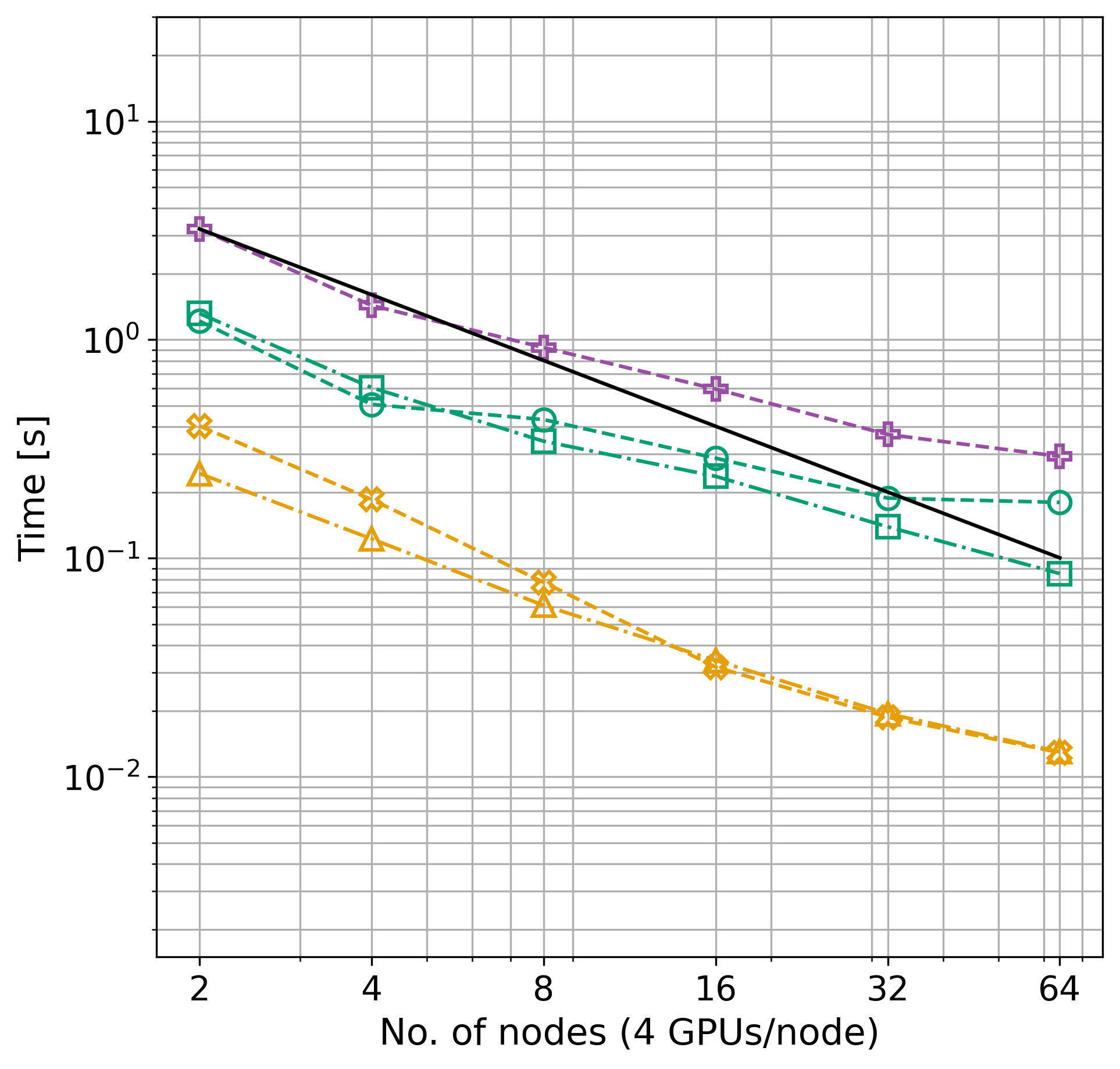}
\end{subfigure}\hfill
\begin{subfigure}{.55\textwidth}
  \centering
  \includegraphics[width=\linewidth]{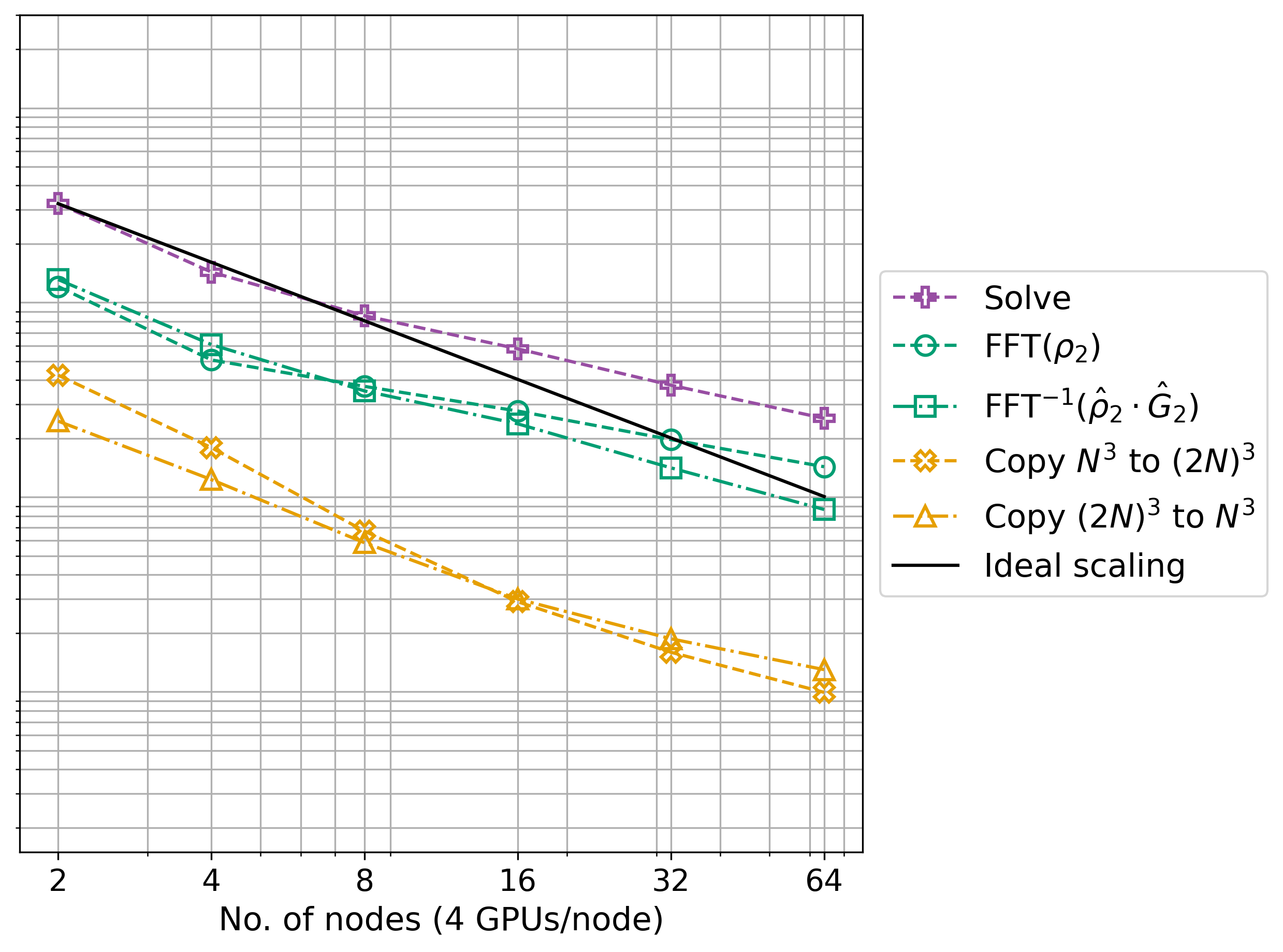}
\end{subfigure}
\caption{Strong scaling results on Alps GPUs for both the Hockney-Eastwood (left) and the modified Vico-Greengard solver (right) with a problem size of $N^3 = 512^3$, using the \texttt{heFFTe} parameters of pencil decomposition, a2av communication, no reordering, and GPU-aware enabled. The efficiency goes down to around $35\%$ for the last grid point.}
\label{512_strong_scaling_daint}
\end{figure}

\begin{figure}[H]
\centering
\begin{subfigure}{.425\textwidth}
  \centering
  \includegraphics[width=\linewidth]{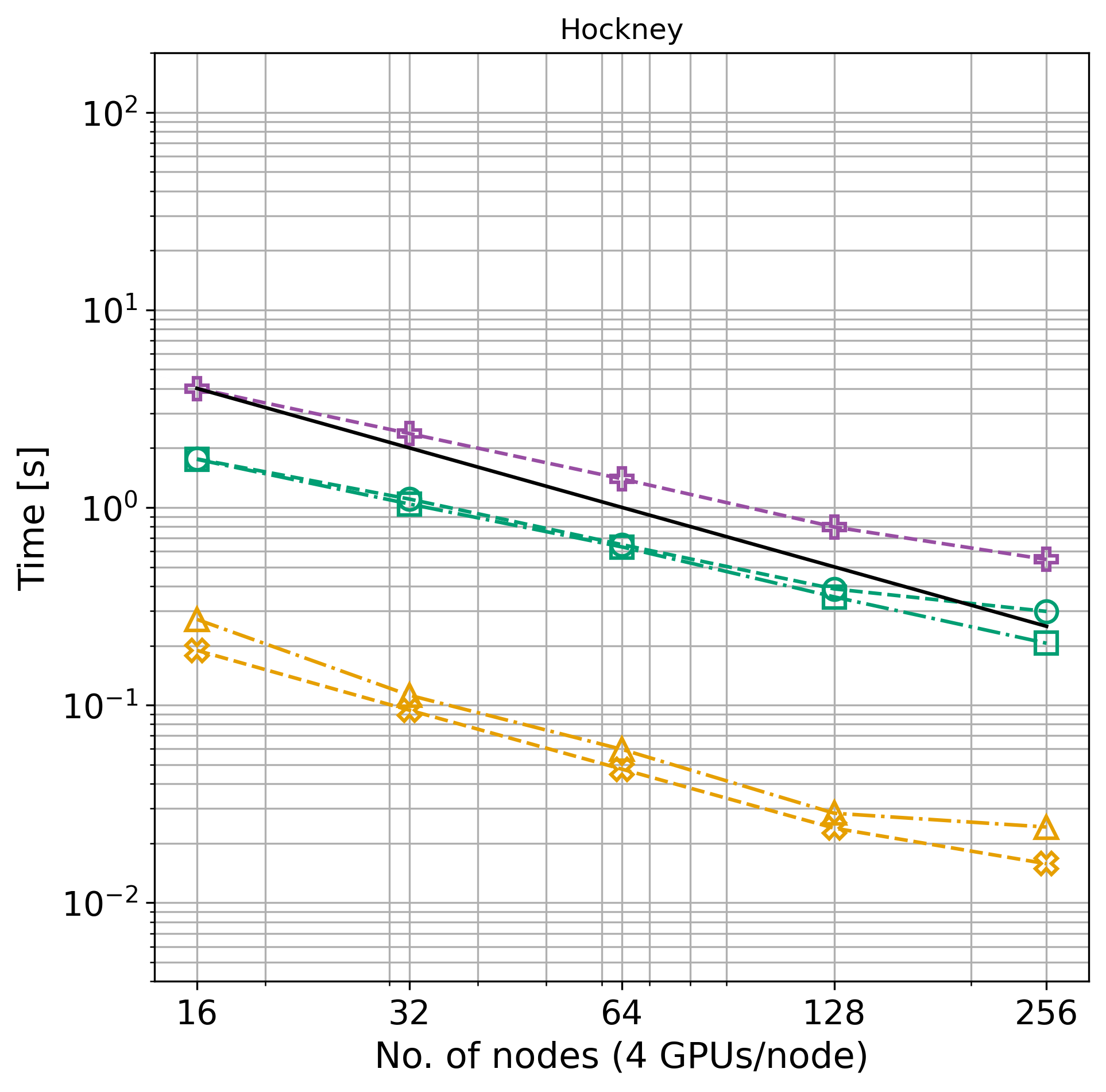}
\end{subfigure}\hfill
\begin{subfigure}{.55\textwidth}
  \centering
  \includegraphics[width=\linewidth]{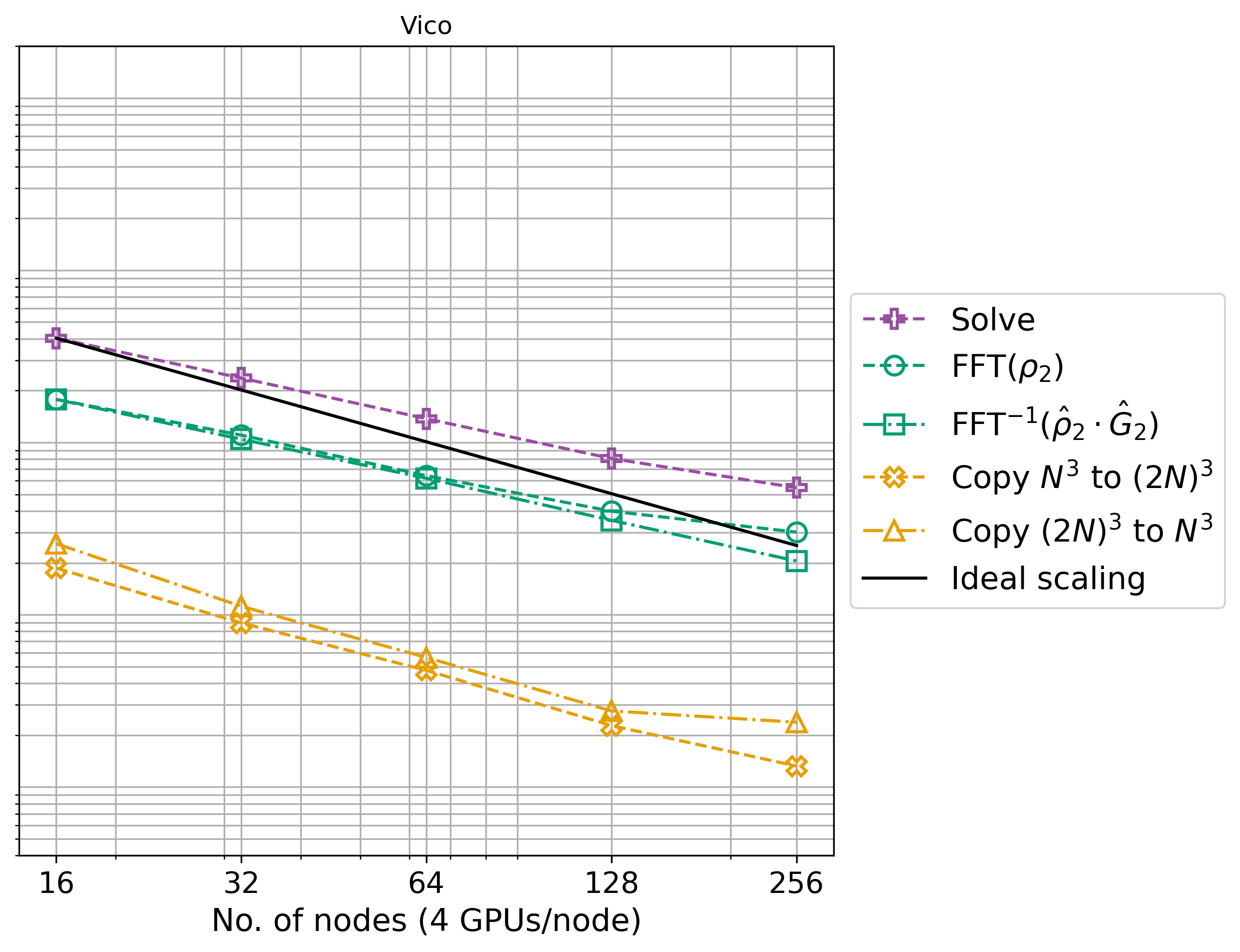}
\end{subfigure}
\caption{Strong scaling results on Alps GPUs for both the Hockney (left) and the modified Vico solver (right) with a problem size of $N^3 = 1024^3$, using the \texttt{heFFTe} parameters of pencil decomposition, a2av communication, no reordering, and with GPU-aware enabled. The efficiency goes down to around $45\%$ for the last grid point.}
\label{1024_strong_scaling_daint}
\end{figure}

\begin{figure}[H]
\centering
\begin{subfigure}{.425\textwidth}
  \centering
  \includegraphics[width=\linewidth]{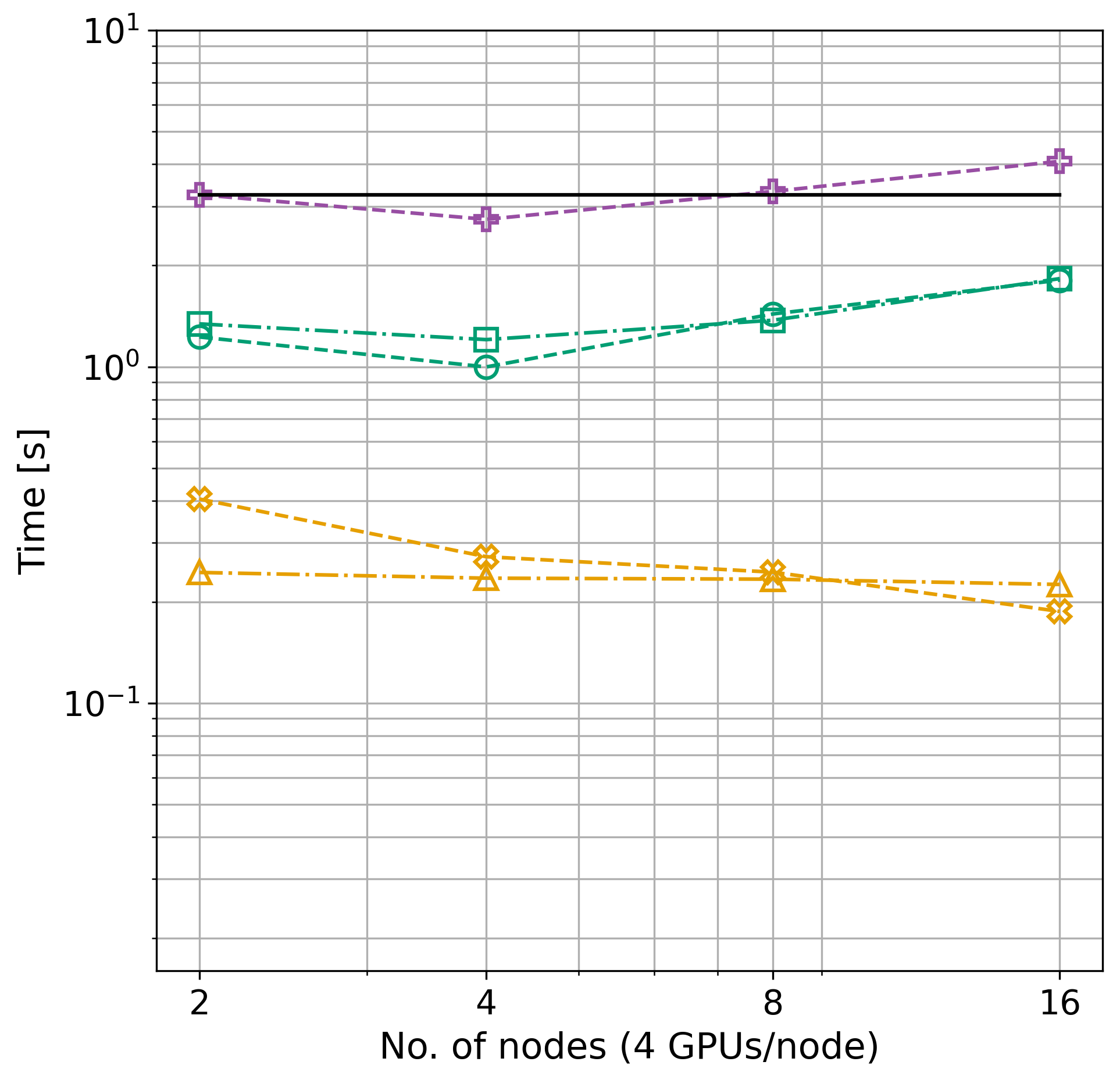}
\end{subfigure}\hfill
\begin{subfigure}{.55\textwidth}
  \centering
  \includegraphics[width=\linewidth]{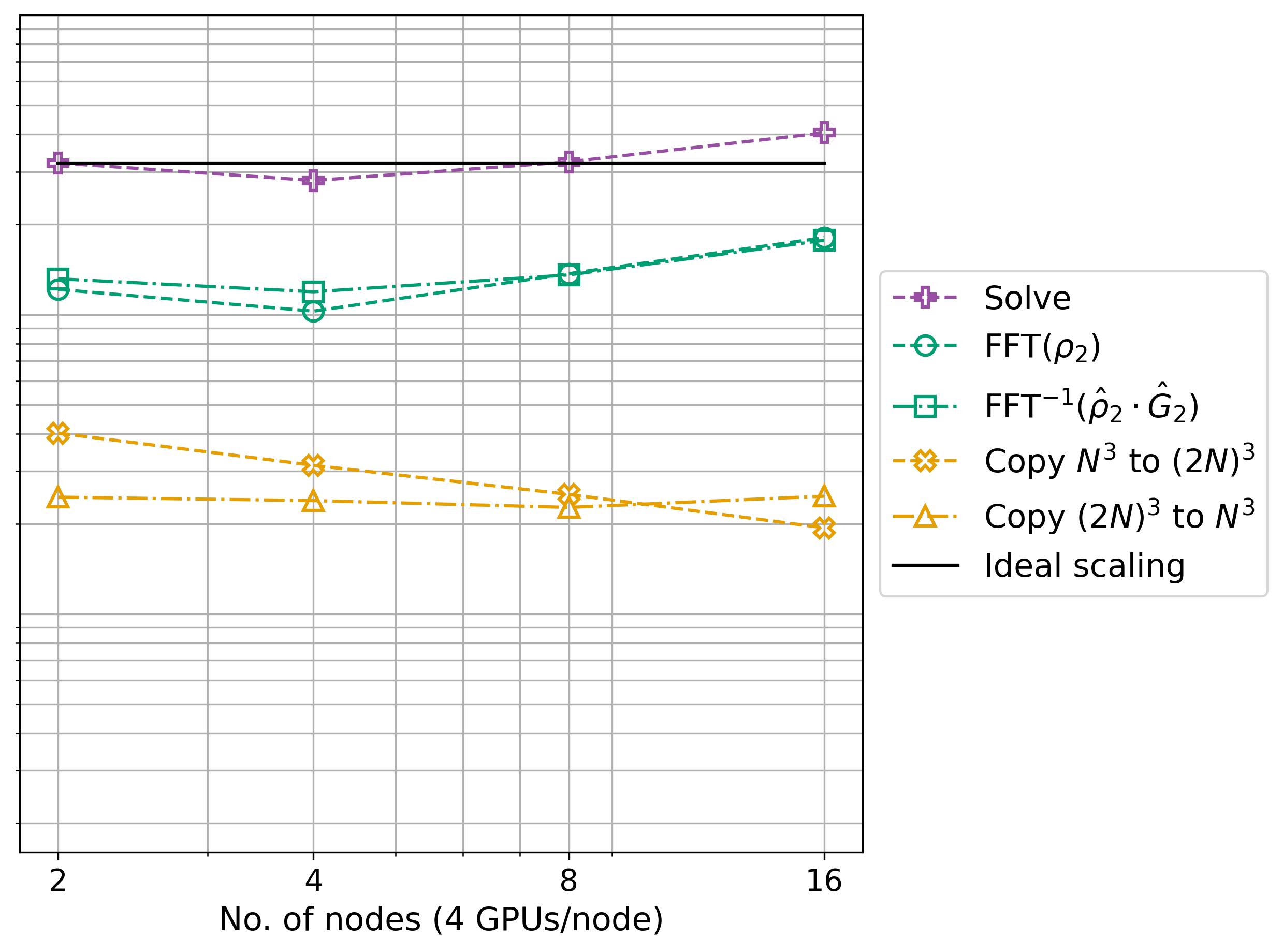}
\end{subfigure}
\caption{Weak scaling results on Alps GPUs for both the Hockney-Eastwood (left) and the modified Vico-Greengard solver (right), starting from a problem size of $N^3 = 512^3$ and increasing the workload proportionally to the node increase until reaching a problem size of $N^3 = 1024^3$. The \texttt{heFFTe} parameters used are pencil decomposition, a2av communication, no reordering, and GPU-aware enabled. The efficiency goes down to around $80\%$.}
\label{weak_scaling_daint}
\end{figure}

On Alps, we compile IPPL using Cuda 12.4 and Cray MPICH 8.1.30, and the same versions of Kokkos and Heffte as for the Perlmutter case (Kokkos 4.1.0 \& Heffte 2.4.0). We first of all make the observation that, in terms of absolute timings, the solver is about twice as fast on the GH200 as on the A100 GPUs. We observe similar scaling behaviour as on Perlmutter, with the difference that on the GH200 nodes the FFT library stops scaling much sooner than on the A100 GPUs. For example, in the $512^3$ strong scaling case (see \cref{512_strong_scaling_daint}), for both the Hockney-Eastwood and the Vico-Greengard case, we stay above $50\%$ efficiency up to 32 nodes (128 GPUs), but if we scale this problem size further it deteriorates, reaching efficiencies around $35\%$ for 64 nodes. The same is seen in \cref{1024_strong_scaling_daint}, where efficiency stays above $60\%$ up to 128 nodes, but goes down to $\approx 45\%$ at 256 nodes. We have observed this scaling behaviour when benchmarking the heFFTe library on Alps. Future versions of heFFTe may have better performance on the GH200 chips \cite{stoyanov_private_2024}. Weak scaling efficiency (\cref{weak_scaling_daint}) stays around $80\%$ up to 16 nodes, and, similarly to Perlmutter, shows the trend of the FFT affecting parallel efficiency due to the communication cost which increases with node count.

\subsubsection{Lumi}

\begin{figure}[H]
\centering
\begin{subfigure}{.425\textwidth}
  \centering
  \includegraphics[width=\linewidth]{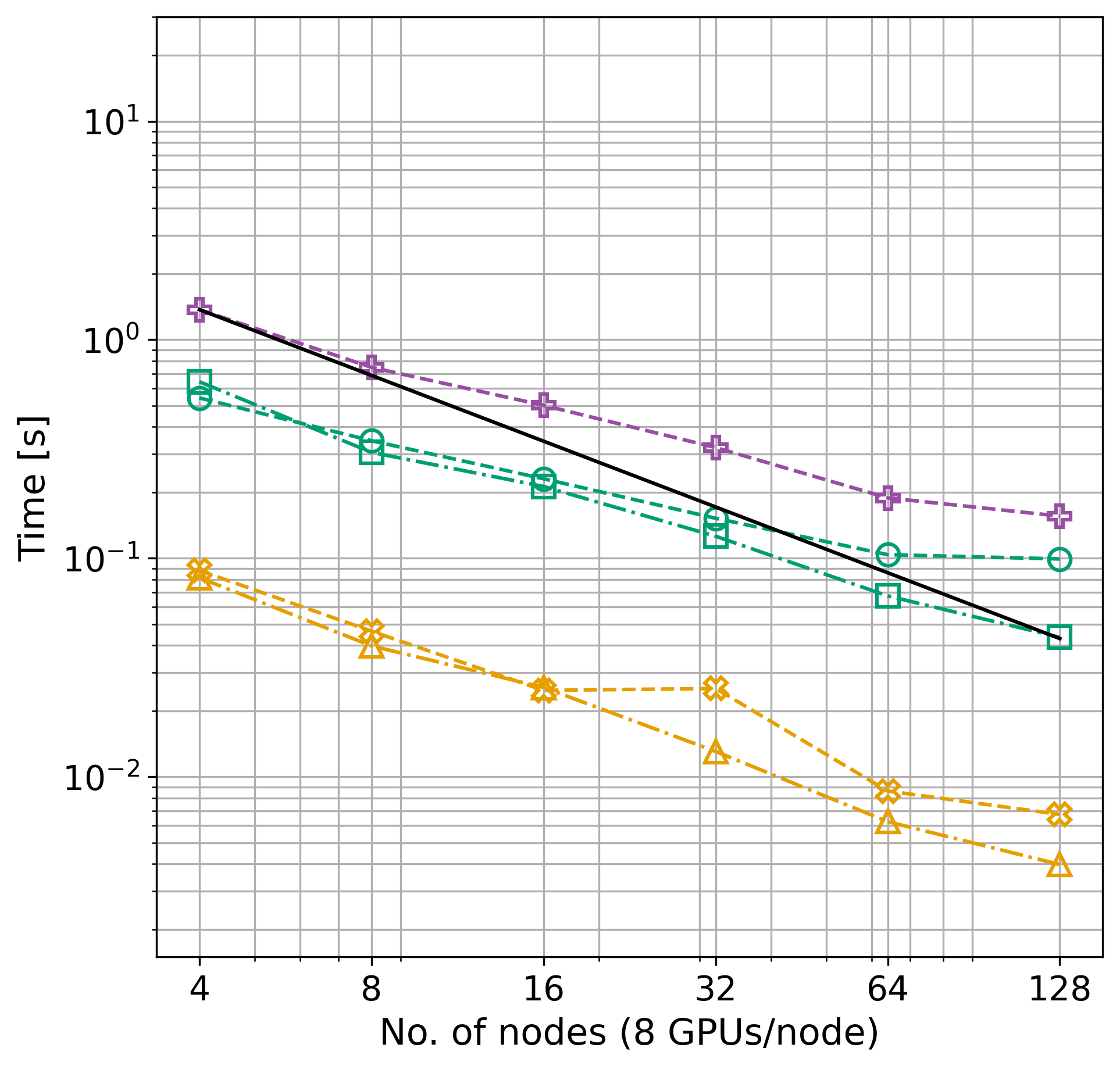}
\end{subfigure}\hfill
\begin{subfigure}{.55\textwidth}
  \centering
  \includegraphics[width=\linewidth]{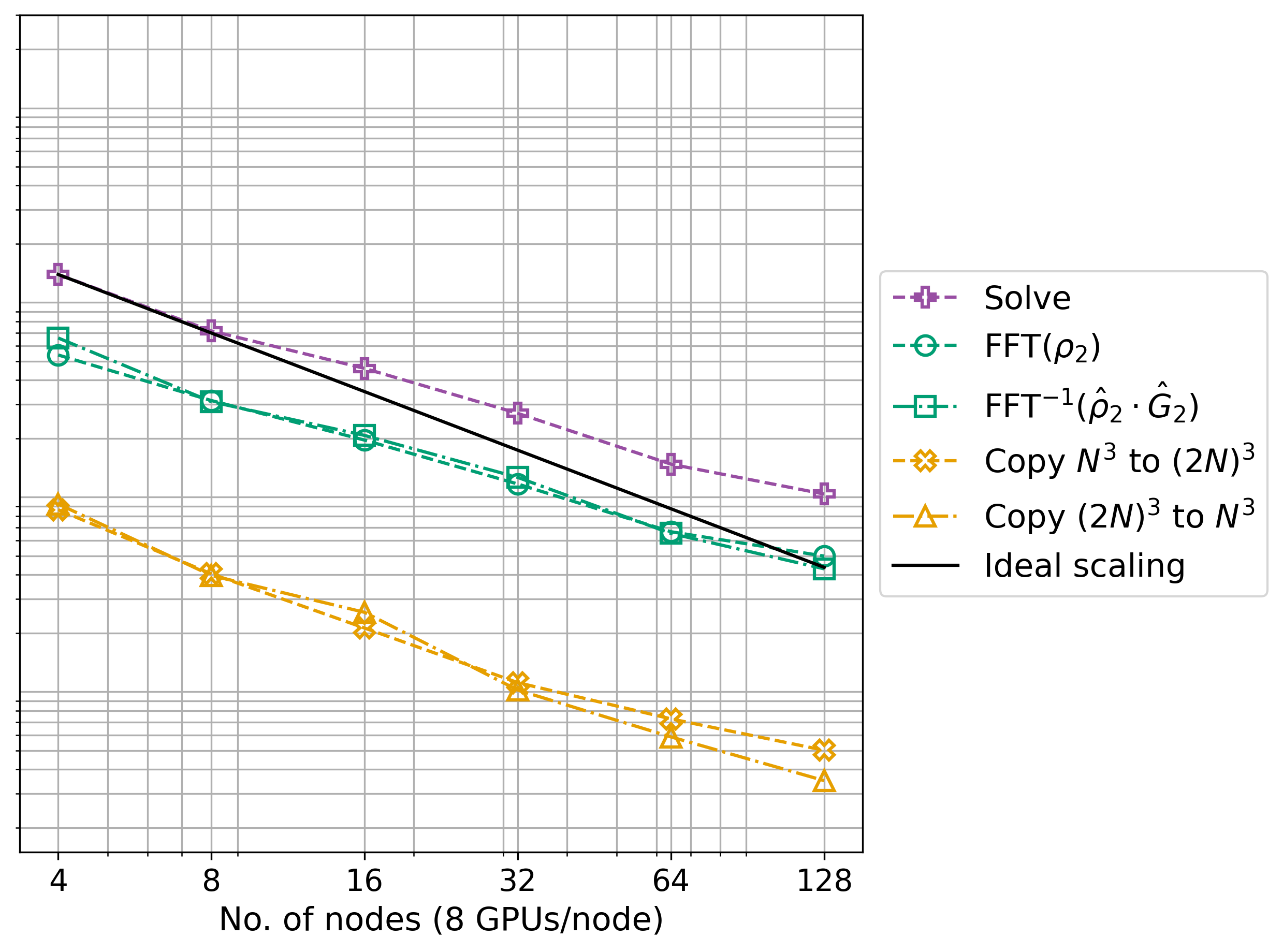}
\end{subfigure}
\caption{Strong scaling results on Lumi GPUs for both the Hockney-Eastwood (left) and the modified Vico-Greengard solver (right) with a problem size of $N^3 = 512^3$, using the \texttt{heFFTe} parameters of pencil decomposition, a2av communication, no reordering, and GPU-aware enabled. For the last grid point, the efficiency drops to around $27\%$ for the Hockney-Eastwood case, and to $41\%$ in the modified Vico-Greengard case.}
\label{512_strong_scaling_lumi}
\end{figure}

\begin{figure}[H]
\centering
\begin{subfigure}{.425\textwidth}
  \centering
  \includegraphics[width=\linewidth]{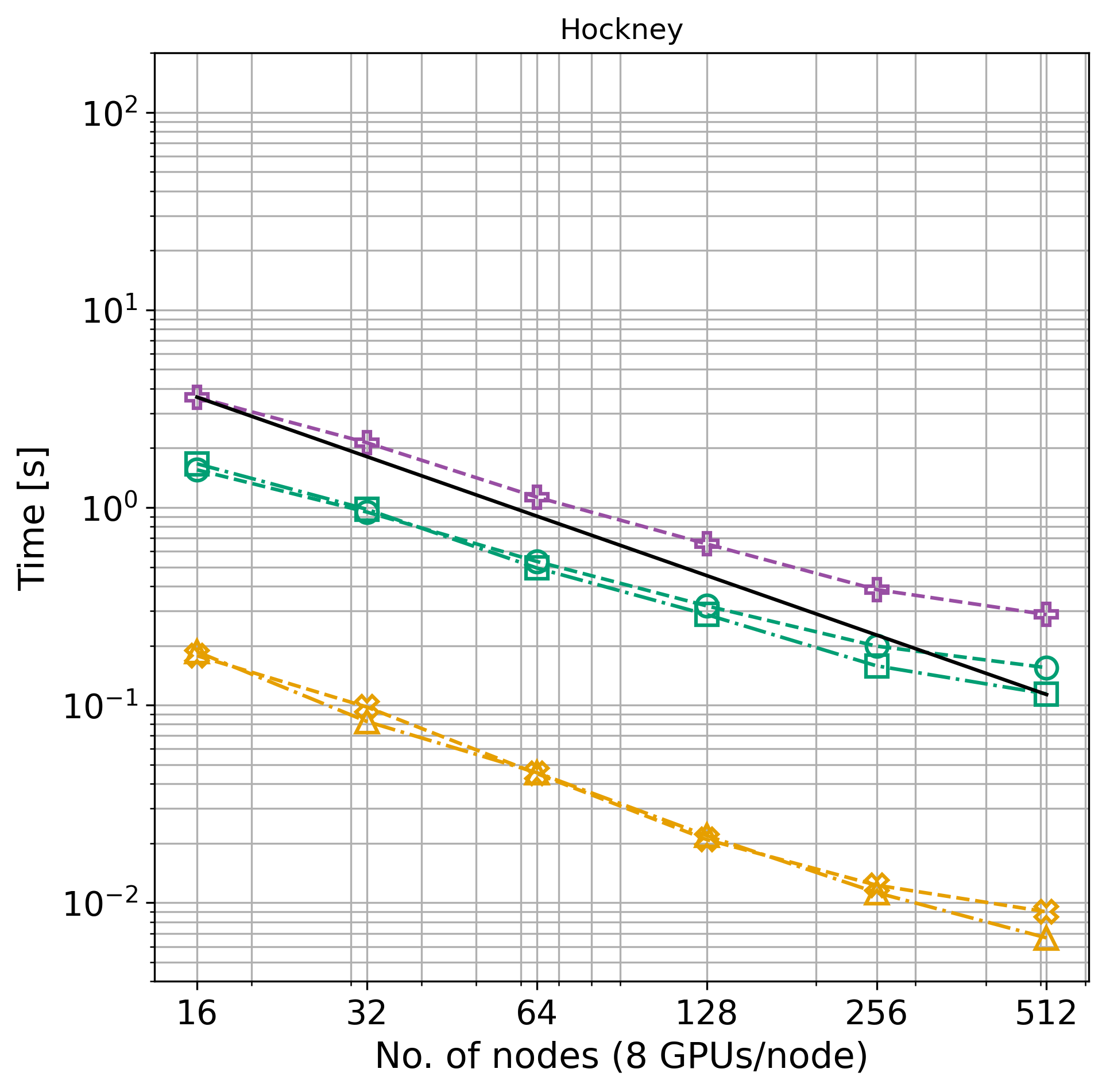}
\end{subfigure}\hfill
\begin{subfigure}{.55\textwidth}
  \centering
  \includegraphics[width=\linewidth]{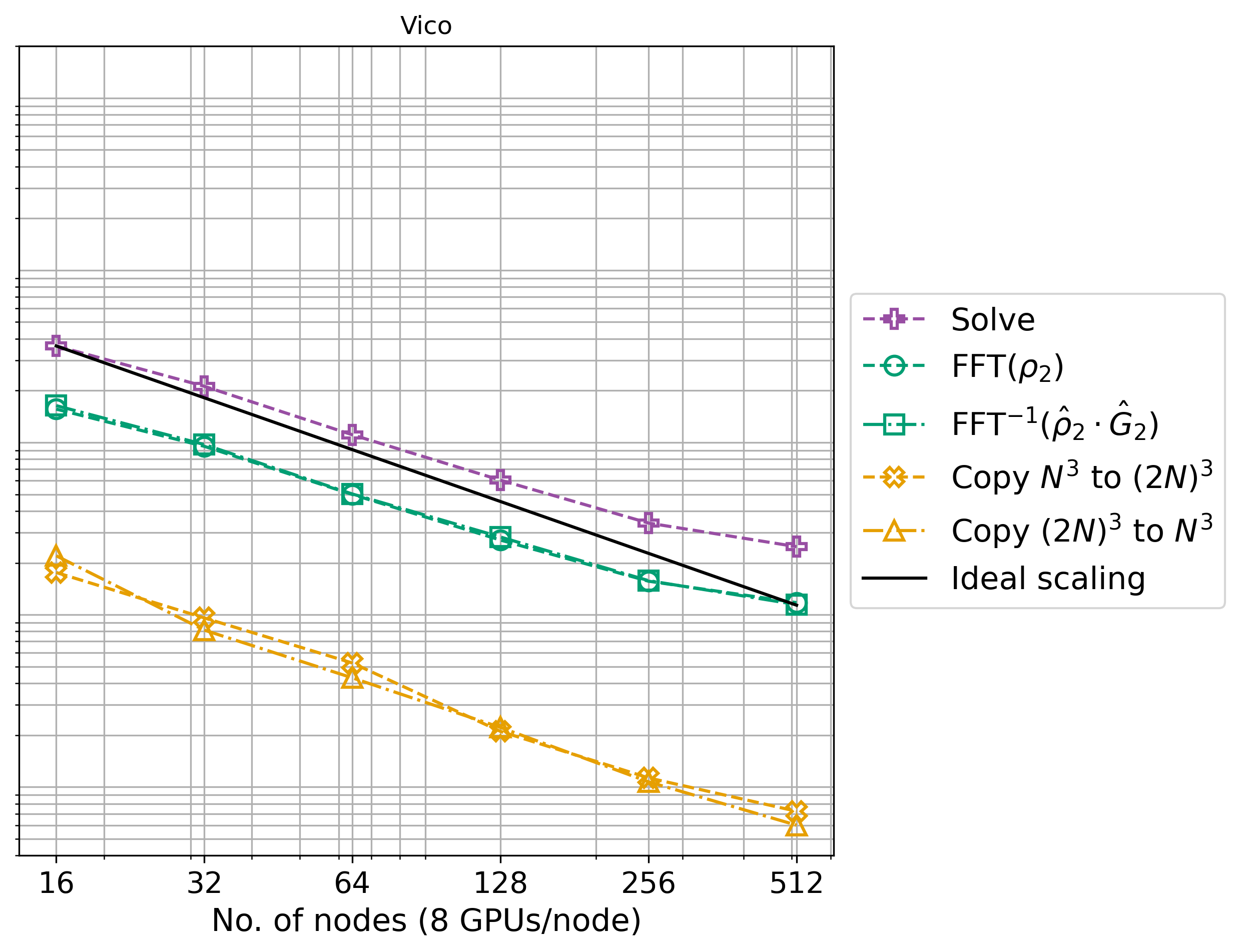}
\end{subfigure}
\caption{Strong scaling results on Lumi GPUs for both the Hockney (left) and the modified Vico solver (right) with a problem size of $N^3 = 1024^3$, using the \texttt{heFFTe} parameters of pencil decomposition, a2av communication, no reordering, and with GPU-aware enabled. The efficiency drops to $39\%$ for the Hockney-Eastwood case, and stays above $45\%$ for the modified Vico-Greengard case.}
\label{1024_strong_scaling_lumi}
\end{figure}

\begin{figure}[H]
\centering
\begin{subfigure}{.425\textwidth}
  \centering
  \includegraphics[width=\linewidth]{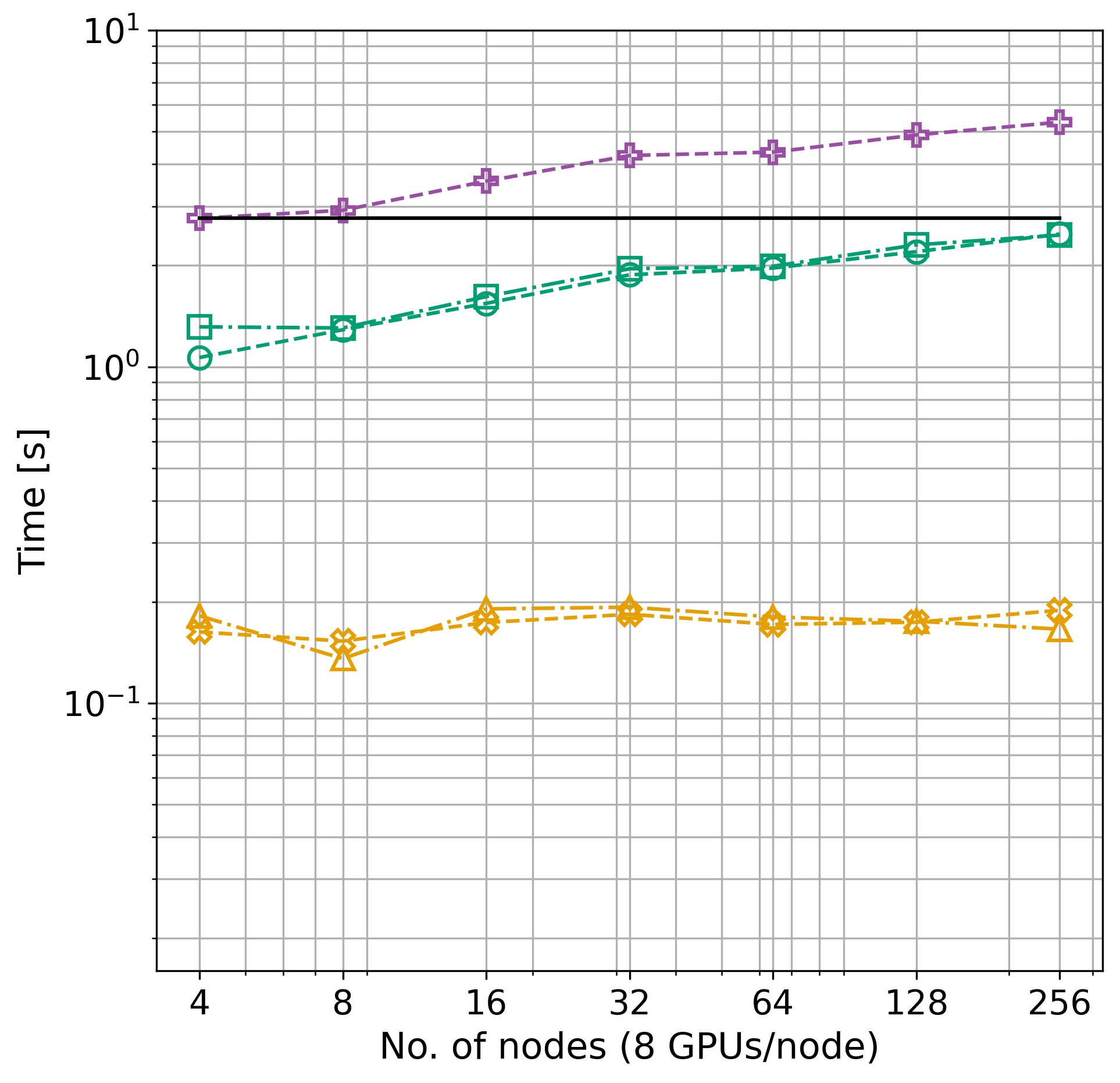}
\end{subfigure}\hfill
\begin{subfigure}{.55\textwidth}
  \centering
  \includegraphics[width=\linewidth]{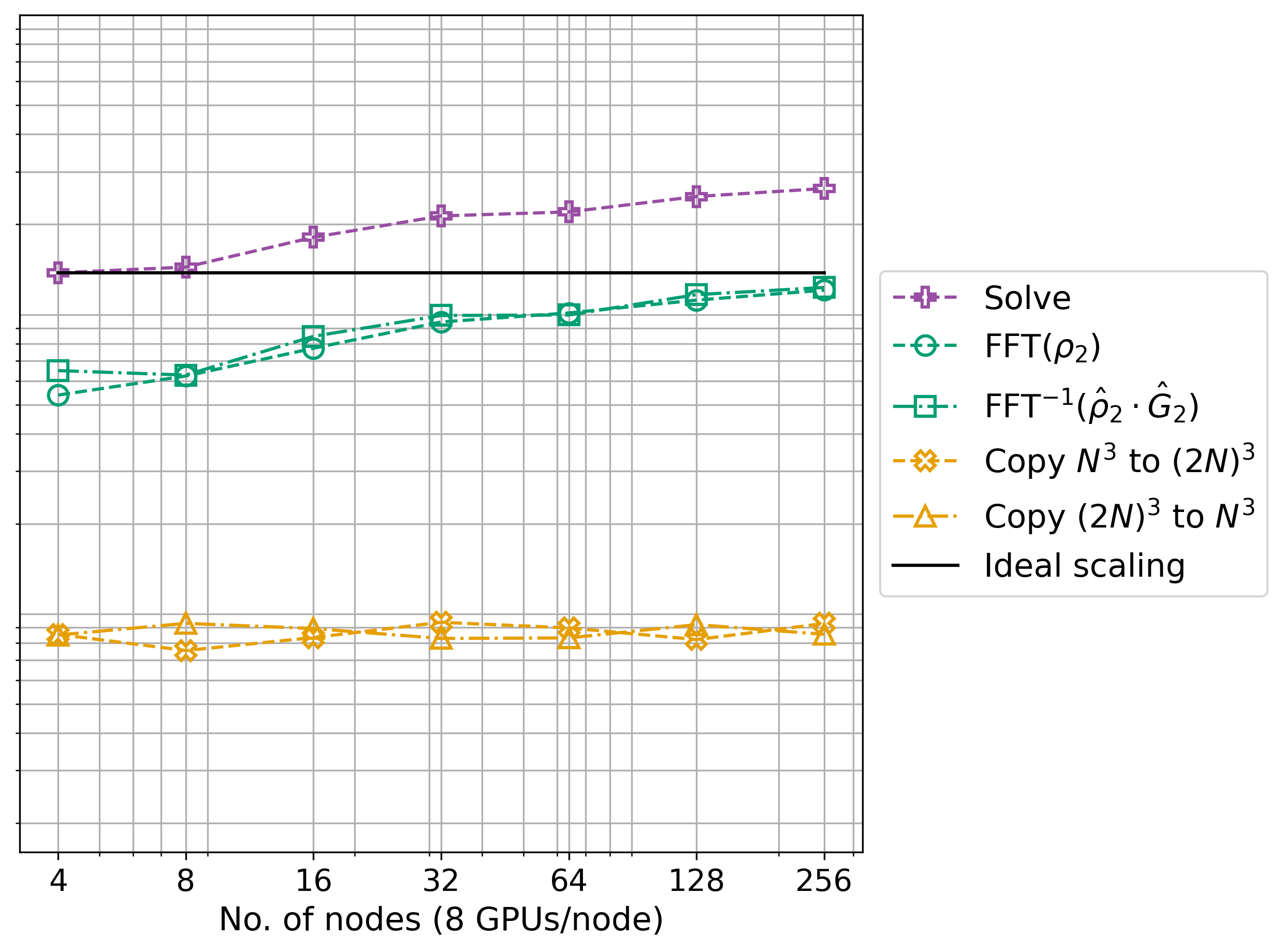}
\end{subfigure}
\caption{Weak scaling results on Lumi GPUs for both the Hockney-Eastwood (left) and the modified Vico-Greengard solver (right), starting from a problem size of $N^3 = 512^3$ and increasing the workload proportionally to the node increase until reaching a problem size of $N^3 = 1024^3$. The \texttt{heFFTe} parameters used are pencil decomposition, a2av communication, no reordering, and GPU-aware enabled. The efficiency stays above $50\%$.}
\label{weak_scaling_lumi}
\end{figure}

On Lumi, we compile IPPL using Rocm 6.0.3 and Cray MPICH 8.1.29, and Kokkos 4.5.0\footnote{This version differs from the ones used for the Nvidia architecture as we faced some compatibility issues between the AMD architecture on Lumi and Kokkos 4.1.0.} \& Heffte 2.4.0. In Figure \ref{512_strong_scaling_lumi}, we see that the overall scaling for the $512^3$ case is worse on this AMD architecture than on the Nvidia A100s, but better than on the Grace-Hopper nodes (Alps). Furthermore, here we observe that the modified Vico-Greengard method scales slightly better than Hockney-Eastwood: the efficiency dips below $60\%$ after we reach a node count of 32 nodes in the Hockney-Eastwood case, but in the Vico-Greengard case it starts to dip below $60\%$ after reaching a node count of 64. For the bigger problem size of $1024^3$ (see Figure \ref{1024_strong_scaling_lumi}), both solvers show the same scaling behaviour, with again the modified Vico-Greengard solver showing a slightly better efficiency. The scalings are also similar to Alps in Figure \ref{1024_strong_scaling_daint}. The weak scaling plots in Figure \ref{weak_scaling_lumi} show again the effect of communication costs in the parallel FFT with increasing node counts.
 
\subsection{Memory footprint}

In terms of memory footprint, one can estimate the gain in memory from using the improved Vico-Greengard by looking at the implementation driven memory requirements. In our code, most of the memory footprint comes from the allocation of the fields used in the Poisson solve. To compute the memory requirements for these fields, we multiply the size of the field, i.e. the total number of values they store, with the amount of bytes required for the precision of the data type we use for them (8 bytes for double precision, and 4 bytes for integer precision).

Below we provide a list of the fields for each method, including their data type, and the estimated memory footprint:
\begin{itemize}
    \item Hockney-Eastwood: There are 5 double precision fields of size $(2N)^3$, and 3 integer fields of size $(2N)^3$. There is one double precision field of size $N^3$, as we reuse the $\rho$ field to write the solution $\phi$ in place. The total memory footprint of this algorithm for a problem size $N^3$ is therefore $M_{H} = 5\cdot (2N)^3\cdot 8 + 3 \cdot (2N)^3 \cdot 4 + N^3\cdot 8$.
    \item Vico-Greengard: Except for the 3 integer fields of size $(2N)^3$, it has the same amount of fields as the Hockney-Eastwood method. In addition, we have two double precision fields of size $(4N)^3$. So we have $M_{V} = 5\cdot (2N)^3\cdot 8 + N^3\cdot 8 + 2\cdot (4N)^3 \cdot 8$.
    \item Modified Vico-Greengard: We have the same fields as the Hockney-Eastwood method except the 3 integer fields of size $(2N)^3$. In addition, we have one double precision field of size $(2N+1)^3$, for the DCT. So we have $M_{V\_DCT} = 5\cdot (2N)^3\cdot 8 + N^3\cdot 8 + (2N+1)^3 \cdot 8$.
\end{itemize}

We can already see that, only by accounting for the memory costs of the IPPL fields, the modification of the Vico-Greengard algorithm allows to reduce the prohibitive cost of the original algorithm. For example, in the $1024^3$ case, the original Vico-Greengard method would incur a 1.5 TB footprint only for storing IPPL fields, whereas in the modified case this is reduced to approximately 421 GB.
\setlength{\tabcolsep}{2pt} % Default value: 6pt
\renewcommand{\arraystretch}{1.3} % Default value: 1
\begin{table}[h]
\centering
\resizebox{\columnwidth}{!}{
\begin{tabular}{ |c|c|c|c|c|c| } 
\hline
Method & Size & IPPL theoretical [GB] & heFFTe Overhead [GB] & Total [GB] & Measured Footprint [GB]\\
\hline
Hockney-Eastwood & $512^3$ & 56 & 5 & 61 & 60 \\ 
\cline{2-6}
            & $1024^3$ & 455 & 40 & 495 & 480 \\ 
\hline
Modified Vico-Greengard & $512^3$ & 53 & 30 & 83 & 87 \\ 
\cline{2-6}
           & $1024^3$ & 421 & 240 & 661 & 700 \\ 
\hline
\end{tabular}
}
\caption{Theoretical memory estimate of IPPL fields and heFFTe workspace overhead estimate for the different algorithms and problem sizes studied in the scaling studies. We also provide a comparison of the measured memory (last column) using the Kokkos profiling tools and the total estimated memory footprint (fifth column) of the simulation.}
\label{theoretical_memory}
\end{table}

\begin{figure}[h]
\centering
\begin{subfigure}{.49\textwidth}
  \centering
  \includegraphics[width=\linewidth]{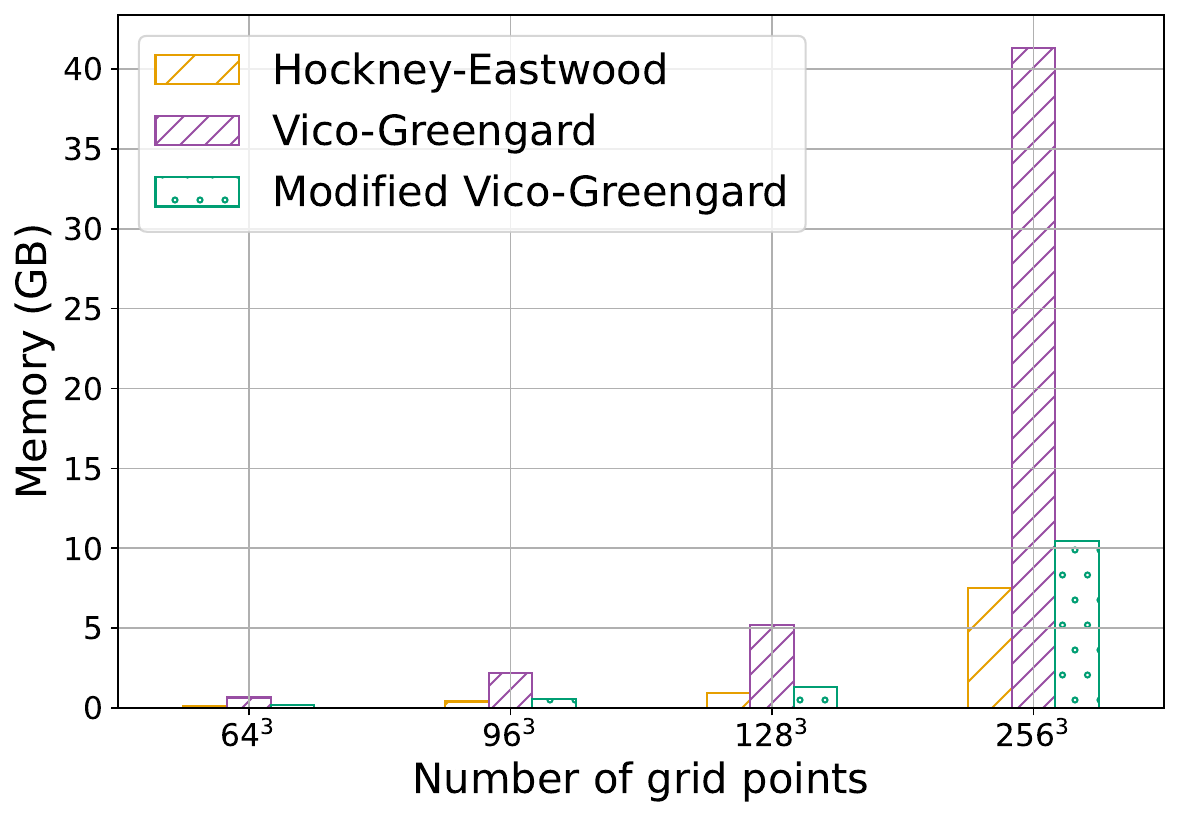}
\end{subfigure}%
\begin{subfigure}{.49\textwidth}
  \centering
  \includegraphics[width=\linewidth]{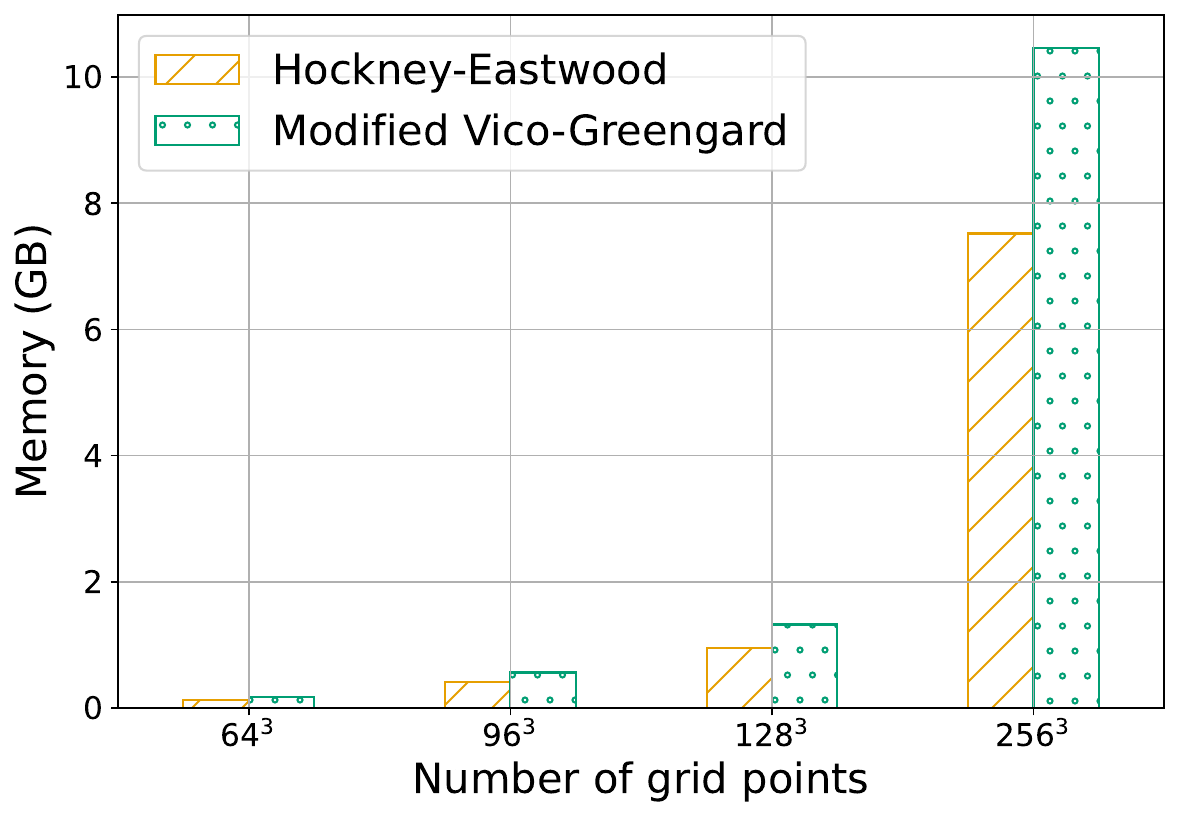}
\end{subfigure}
\caption{Memory footprints of the solvers for problem sizes $64^3, 96^3, 128^3,\text{ and } 256^3$, obtained using Kokkos Highwater on Alps GPU. Left: Memory comparison of Hockney-Eastwood, original Vico-Greengard, and modified Vico-Greengard solvers. Right: Comparison of only Hockney-Eastwood and the modified Vico-Greengard solver, which has an algorithmic optimization.}
\label{memory_plots}
\end{figure}

Apart from the IPPL fields, the other major memory cost in the simulation comes from the heFFTe library, which incurs some overhead as it needs to define a workspace for the FFTs. \Cref{theoretical_memory} shows the theoretical IPPL estimate, the heFFTe overhead estimate, and a comparison with the actual memory footprint (from \cref{memory_table}) for $512^3$ and $1024^3$ problem sizes for both Hockney-Eastwood and modified Vico-Greengard algorithms. Furthermore, we also conducted a Kokkos memory profiling of all three methods for different problem sizes, shown in \cref{memory_plots}. As we can see, with the algorithmic modification of using the discrete cosine transform, the memory footprint of the spectral method is brought down to be similar to the Hockney-Eastwood method. The Vico-Greengard algorithm therefore becomes a viable solver for the free-space Poisson equation in the context of mesh-based methods.

The modified Vico-Greengard method still has a higher memory cost than Hockney-Eastwood for the same number of grid-points due to the higher heFFTe overhead. However, for sufficiently smooth solutions, the total memory required for a similar accuracy of simulation is still less in the modified Vico-Greengard case due to the spectral accuracy, as we explain in the next section.

\subsection{Roofline analysis}

In order to further understand the performance of our code, we also conducted a roofline analysis. Since the solver is dominated by the FFT operation, the roofline analysis is on the heFFTe real-to-complex benchmark. We run the benchmark for two problem sizes, $1024^3$ and $2048^3$, which correspond to problem sizes of $512^3$ and $1024^3$ respectively for the solver, due to the doubling of the domain (see \cref{sec:HockneyEastwood}). We run on different number of GPUs on all architectures: 16, 32, and 64 for $1024^3$, and 64, 128, and 256 for $2048^3$.

\begin{figure}[h]
    \centering
    \includegraphics[width=0.65\linewidth]{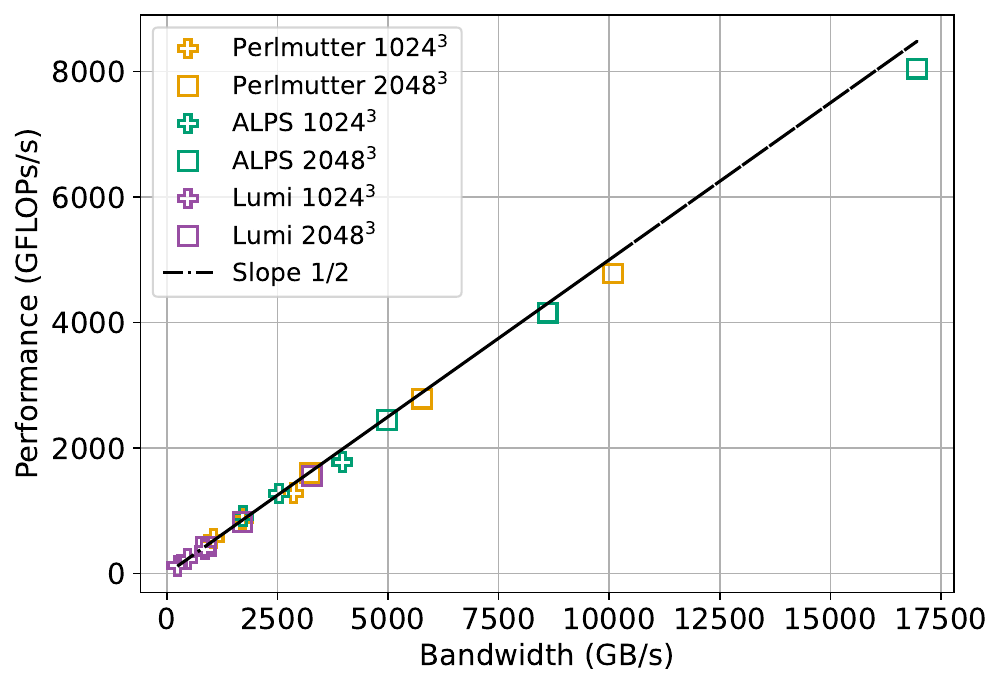}
    \caption{Floating point operations per second vs. the bandwidth used by the program, for Perlmutter (Nvidia A100), ALPS (GH200) and Lumi (AMD MI250X) supercomputers. All the datapoints lie on the dotted line, which has slope 1/2, and corresponds to the arithmetic intensity (FLOPs/byte).}
    \label{flops_byte_plot}
\end{figure}
\begin{figure}[h]
    \centering
    \includegraphics[width=0.7\linewidth]{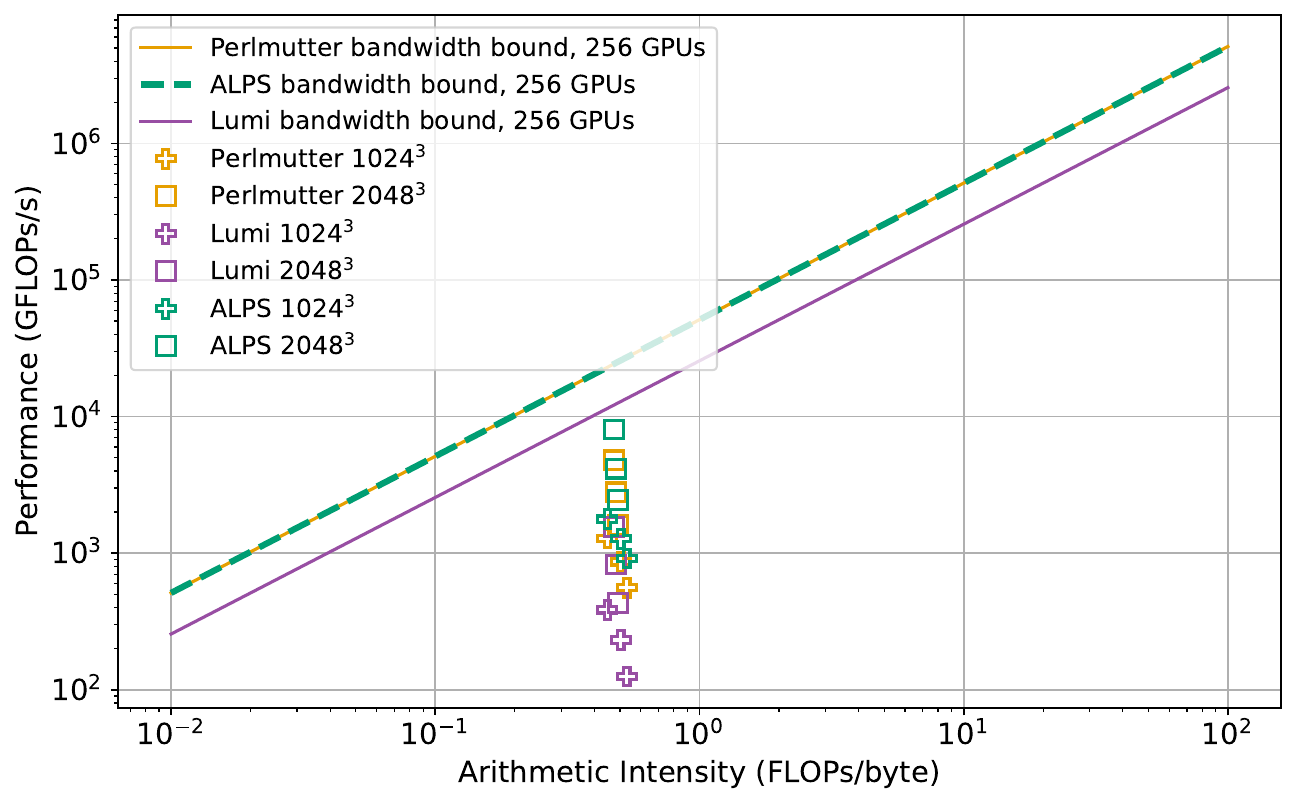}
    \caption{Roofline plot showing the bandwidth bound for 256 GPUs for each machine, as well as the measured points. The bandwidth bound lower GPU counts is not shown for the sake of readability. The peak performance, which is in the petaFLOP regime for these machines, is not shown as it is surpasses the scales shown in the plot. The FFT is clearly memory-bound. }
    \label{roofline_plot}
\end{figure}

\cref{flops_byte_plot} shows the number of floating point operations per second vs. the memory bandwidth of these runs i.e. computation vs. communication. As we can see, the arithmetic intensity (FLOPs/byte) stays in the region 0.44 - 0.54, FLOPs/byte, and is roughly the same for both problem sizes on all architectures. This is as expected: for a problem size $N$, the computational complexity of the FFT is given by $\mathcal{O}(N\text{log}N)$, and the data movement is of the order of $N$, which makes the arithmetic intensity scale as $\text{log}N$. For the problem sizes $1024^3$ and $2028^3$, the increase in arithmetic intensity should therefore be negligible. The roofline plot is shown in \cref{roofline_plot}. The benchmark falls in the memory-bound regime for all architectures, which is as expected for a distributed FFT. The profiling data for all runs can be found in \cref{performance_analytics}.

\subsection{Summary of solvers}

In this section, we juxtapose all three solvers taking into account both the accuracy and the memory footprint. This is because the Vico-Greengard solver has high accuracy and spectral convergence for smooth solutions. Therefore, according to the smoothness of the source distribution in the Poisson equation, one may need less grid points to achieve the same result as the Hockney-Eastwood solver. We show this final comparison in \cref{comparison_table}, where we take the values for the very smooth case of the Gaussian source (as in \cref{gaussian}). As we can see, for this smooth distribution, even the non-modified original Vico-Greengard solver has less memory requirements than the Hockney-Eastwood method for a given accuracy. % as it converges so fast that even to achieve an accuracy of $10^{-9}$ we only need $32^3$ gridpoints, which is less memory intensive than the amount of gridpoints we would need for the second-order Hockney-Eastwood method.
However, an advantage of the modified Vico-Greengard solver, which uses the DCT, is that one could make it the default solver regardless of the smoothness of distribution, since the price to pay in terms of memory is only slightly higher (same order of magnitude) than the Hockney-Eastwood solver. On the other hand, the original Vico-Greengard algorithm is only advantageous for smooth distributions. However, in targeted applications like particle accelerators and plasma physics, the smoothness of the density distribution evolves continuously over time and is not known a priori.

\setlength{\tabcolsep}{10pt} % Default value: 6pt
\renewcommand{\arraystretch}{1.3} % Default value: 1
\begin{table}[h]
\centering
\resizebox{\columnwidth}{!}{
\begin{tabular}{|l|c|c|c|c|c|c|}
\hline
\multicolumn{1}{|c|}{Algorithm} & \multicolumn{2}{c|}{Hockney-Eastwood} & \multicolumn{2}{c|}{Vico-Greengard} & \multicolumn{2}{c|}{Modified Vico-Greengard (DCT)}\\
\cline{1-7}
\multicolumn{1}{|c|}{Accuracy} & Gridsize & Memory [MB] & Gridsize & Memory [MB] & Gridsize & Memory [MB] \\
\hline
 $10^{-1}$ & $8^3$ & $\sim 10^{-1}$ & $8^3$ & $\sim 10^{0}$ & $8^3$ & $\sim 10^{-1}$ \\ 
 $10^{-2}$ & $32^3$ & $\sim 10^{1}$ & -  & - & - & - \\ 
 $10^{-3}$ & $128^3$ & $\sim 10^{3}$ & $16^3$ & $\sim 10^{1}$ & $16^3$ & $\sim 10^{0}$ \\ 
 $10^{-5}$ & $512^3$ & $\sim 10^{4}$ & - & - & - & - \\ 
 $10^{-9}$ & $\approx (10^4)^3$ & $\sim 10^{8}$ & $32^3$ & $\sim 10^{2}$ & $32^3$ & $\sim 10^{1}$\\ 
\hline
\end{tabular}
}
\caption{Number of gridpoints and estimated memory footprint for all three solvers according to the level of accuracy desired (where the accuracy denotes the relative error between computed and exact solution). The number of gridpoints and accuracy values are taken from the Gaussian correctness test (see \cref{gaussian}), which is a very smooth distribution.}
\label{comparison_table}
\end{table}

\section{Conclusion}

In this work, we introduce an enhancement to the fast free space Poisson solver initially proposed by Vico et al. in \cite{vico_fast_2016}. Our improvement reduces the memory usage of the algorithm to just one-eighth of its original amount, bringing its memory footprint into the same order-of-magnitude range as the Hockney-Eastwood solver \cite{hockney_computer_1988}, the current state-of-the-art method for addressing the free space Poisson problem in particle-mesh simulations. The new solver offers a significant advantage by providing spectral accuracy for smooth source terms. As a result, it needs fewer grid points compared to the state-of-the-art second-order Hockney-Eastwood solver to achieve the same level of accuracy for sufficiently smooth distributions. This makes it an excellent choice for overcoming memory and runtime constraints in large-scale, high-resolution simulations.

We have implemented this newly improved solver in a massively parallel and portable framework targeted to exascale machines. We quantify the correctness and accuracy improvements with respect to the Hockney-Eastwood method, as well as the memory improvements with respect to the original algorithm in \cite{vico_fast_2016}. Finally, we demonstrate the efficiency of the solver by running strong and weak scaling analyses on Perlmutter, for both CPU and GPU, using up to $\sim70\%$ of the full machine in the latter case. The results of the strong scaling study show that the efficiency stays above $50\%$ for the problem size of $512^3$, and above $65\%$ in the case of the larger problem size $1024^3$. When running on the Nvidia Grace-Hopper chips on Alps, we see a factor of two improvement in runtime compared to Perlmutter, whereas when running on the AMD MI250X GPUs on Lumi, we get about a factor of $\sim1.5$ improvement. On Alps, the strong scaling study shows a parallel efficiency above $50\%$ for a problem size of $512^3$ up to 32 nodes, and above $60\%$ up to 128 nodes for $1024^3$, after which it starts to deteriorate. On Lumi, the strong scaling study shows a parallel efficiency above $60\%$ up to 32 nodes for a $512^3$ problem size, and up to 128 nodes for the $1024^3$ problem size. Note that the effective number of GPUs per node is twice as much on Lumi than on Alps (8 GPUs per node vs. 4 GPUs per node respectively).
%This is due to the FFTs not scaling well on this architecture for this problem size, and further releases of the heFFTe library may improve on this.

Currently, the new solver has been verified in the context of a plasma physics application, a charge-neutral Penning trap, the results of which can be found in \cref{penning_appendix}. 

In future work, we will apply this free space solver in the context of beam dynamics simulations using OPAL \cite{adelmann_opal_2019}, a particle accelerator modelling code, which is currently being interfaced with the performance portable IPPL framework. 

\section*{Availability}
IPPL is an open source project.\ The source code can be found here: \url{https://github.com/IPPL-framework/ippl}, with the version of IPPL used for this study tagged \texttt{scaling\_study\_vico\_paper}.\ The versions of the build dependencies used on the on each system are listed in the Results section of the paper. The convergence study are found in \texttt{test/solver/TestGaussian\_convergence.cpp}, and the scaling studies were done with \texttt{test/solver/TestGaussian.cpp}. The solver type can be chosen by passing either \texttt{HOCKNEY}, \texttt{VICO}, or \texttt{VICO\_2}, which correspond to the Hockney-Eastwood, Vico-Greengard, and modified Vico-Greengard methods respectively.

\section*{Acknowledgements}
This research used resources of the National Energy Research Scientific Computing Center (NERSC), a Department of Energy Office of Science User Facility using NERSC award ERCAP-M888. We acknowledge access to Alps at the Swiss National Supercomputing Centre, Switzerland under the Paul Scherrer Institute's share with the project ID psi07 and the PASC project with ID c41. We acknowledge CSCS, Switzerland for awarding this project access to the LUMI
supercomputer, owned by the EuroHPC Joint Undertaking, hosted by CSC (Finland)
and the LUMI consortium through CSCS, Switzerland. The Gwendolen cluster at PSI was essential to the successful development of the presented method. We particularly acknowledge the invaluable technical advice provided by the PSI High-Performance Computing and Engineering (HPCE) group, especially Marc Caubet Serrabou. We would also like to acknowledge Stanimire Tomov and Hartwig Anzt from the Innovative Computing Laboratory - University of Tennessee and Miroslav Stoyanov from Oak Ridge National Laboratory for their help with introducing the Discrete Cosine Transform of Type 1 in heFFTe, as well as insights into the heFFTe benchmarks.

 \bibliographystyle{ieeetr} 
 \bibliography{bibliography}

\appendix

\section{Implementation details for copy routines} \label{MPI_implementation}

The purpose of this appendix is to explain the algorithm that we use to do all the copy operations involving moving from the $N^3$ to the $(2N)^3$ grid and vice-versa in an efficient manner such as to not hinder the performance of our solver.

The problem setup is the following: if we have a computational grid of $N^3$ grid points, and we run on $n$ number of MPI ranks, the computational domain is decomposed and distributed among these $n$ ranks. Each of these $n$ ranks then contain only a portion of the fields on the computational domain, corresponding to the sub-domain they hold.

When we then zero-pad to a $(2N)^3$ grid to follow the solver algorithm explained in \cref{sec:HockneyEastwood}, this doubled domain is also domain decomposed among the same $n$ ranks. Therefore, the sizes and extents of the sub-domains of the $N^3$ fields and the $(2N)^3$ fields are not the same in each rank, meaning that communication is required when we want to zero-pad it to get a $(2N)^3$ field or restrict it from the $(2N)^3$ grid to the $N^3$ grid.

\begin{figure}[ht]
    \centering
    \includegraphics[width=7cm]{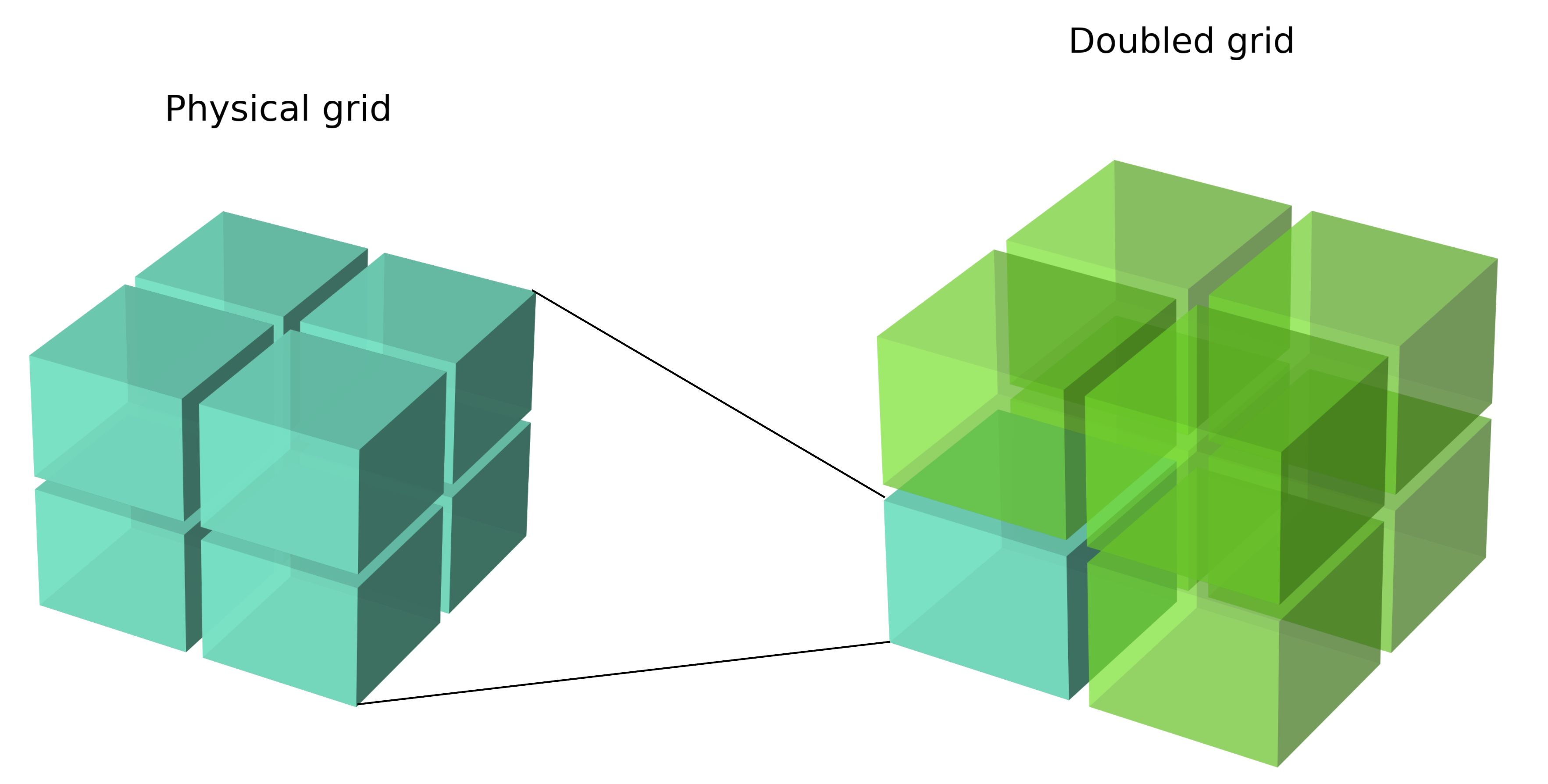}
    \caption{Schematic of communication needed to go from the physical grid ($N_x \times N_y \times N_y$) to the doubled grid ($2N_x \times 2N_y \times 2N_z$). On the right, the blue cube indicates the physical sub-domain of the doubled grid, whereas the green cubes are those where $\rho$ will need to be zero-padded. The cubes themselves represent the domain decomposition of the fields among ranks.}
    \label{phys_to_doub}
\end{figure}

For example, when writing $\rho$ to $\rho_2$, we need communication to get the sub-domains of the field $\rho$ to the rank which corresponds to the same sub-domain in $\rho_2$. A diagram of the problem is shown in Figure \ref{phys_to_doub}.
The implementation of this assignment of a field from the physical grid to a doubled grid is non-trivial. The approach followed is shown in Algorithm \ref{assignment}. All ranks will encounter this algorithm and execute it, such that the end result will be what is desired (Figure \ref{phys_to_doub}). A similar algorithm is used for the restriction from the doubled grid to the physical grid ($\phi_2\rightarrow\phi$). The same applies to the restriction of the Green's function from the $(4N)^3$ or $(2N+1)^3$ grids to the $(2N)^3$ grid, in the case of the Vico-Greengard method and its modified version.

%Furthermore, in this case, since the Green's function is defined in Fourier space first and then an inverse transform is done before the restriction, there is an additional step of reordering the field such that the inverse transform is correct in configuration space. This is because of the way the wavevectors are ordered in heffte (following the standard in-order output as in fftw). When we define the G_L function for the Vico method, we define it in Fourier space following an intuitive ordering (most negative to most positive frequency). Then, when we inverse transform and restrict to the doubled grid, we need to reorder the output of the inverse transform for it to make sense in config space, since the heffte transform assumes the following wavevector ordering: first all positive frequencies, then negative frequencies in descending order. 

\begin{algorithm}[H]
\caption{Assignment from physical to doubled grid. The bounds of the sub-domains contained in each rank are globally known.}\label{assignment}
\begin{algorithmic}

\\

\State $n$ $\gets$ total number of ranks
\State $myRank$ $\gets$ ID of this rank
\State $domains1$ $\gets$ bounds of sub-domains of physical grid
\State $domains2$ $\gets$ bounds of sub-domains of doubled grid

\\

\For{$i\gets 1, n$}
\If{$domains2[i]$ touches $domains1[myRank]$}
    \State intersection $\gets$ intersect $domains2[i]$ with $domains1[myRank]$
    \State Pack $domains1[myRank][intersection]$.
    \State Send packed data to rank $i$.
\EndIf
\EndFor

\\

\For{$i\gets 1, n$}
\If{$domains1[i]$ touches $domains2[myRank]$}
    \State $intersection$ $\gets$ intersect $domains1[i]$ with $domains2[myRank]$
    \State Receive data from rank $i$.
    \State Unpack data into $domains2[myRank][intersection]$.
\EndIf
\EndFor

\end{algorithmic}
\end{algorithm}

\section{Penning trap simulation using the Particle-In-Cell scheme} \label{penning_appendix}
As an application of the Vico-Greengard solver, we use it in combination with the Particle-In-Cell (PIC) scheme for an electrostatic plasma physics simulation. More concretely, we simulate the dynamics of an electron bunch in a Penning trap with external electric and magnetic fields to confine them. First, we introduce the theory behind the PIC method, and then we follow-up with the results, where we compare the simulation using the Vico-Greengard solver with the Hockney-Eastwood solver.

\subsection{The Particle-In-Cell (PIC) scheme} \label{theory-pic}

The particle-in-cell method combines particle tracking in phase space with a mesh-based approach to speed up force calculations. To reduce the computational cost needed to model physical phenomena, a cloud of plasma particles is replaced by a macro-particle with the same charge-to-mass ratio. In the case considered here, these macro-particles evolve in phase space over time according to the Vlasov-Poisson equation and Newton's second law of motion. In order to compute the force fields, the charge density of the particles is computed on the grid points of a fixed mesh (\textbf{scatter}), on which we solve for the electrostatic fields (\textbf{field solve}). These are then interpolated to the macro-particle locations (\textbf{gather}) in order to compute the force which will drive the particles to their new position at the next time-step (\textbf{particle push}) \cite{saez_particle--cell_2011}. The PIC loop consists of the scatter, solve, gather, and push, and is repeated at each time-step until the end of the simulation, schematically shown in \cref{pic_loop}.

\begin{figure}[h]
    \centering
    \resizebox{11cm}{!}{
    \begin{tikzpicture}[
        node distance=4ex and 0em,
        block/.style={rectangle, draw=blue!60, fill=blue!5, 
        text width=9em, text centered, rounded corners, minimum height=3em},
        block_n/.style={rectangle, draw=blue!60, fill=blue!5, 
        text width=10em, text centered, rounded corners, minimum height=3em},
        block1/.style={rectangle, draw=green!60, fill=green!5, 
        text width=9em, text centered, rounded corners, minimum height=3em},
        line/.style={draw, -latex},
        ]
        \node [block] (1) {\textbf{SOLVE:} \\ Calculate field $\phi$ and $\vec{E}$ from $\rho$};
        \node [block, below right= of 1] (2) { \textbf{GATHER:} \\ Interpolate fields from grid to particles};
        \node [block, below left= of 2] (3) {\textbf{PUSH:} \\ Update particle position and velocity};
        \node [block_n, above left= of 3] (4) {\textbf{SCATTER:} \\ Interpolate charge of macro-particles onto grid to obtain $\rho$};
        \node [block1, left=1cm of 4] (5) {\textbf{INITIALIZATION:} \\ Initialize particle positions, velocities, and charges};

        \path [line] (5.east) to[out=0, in=180] (4.west);
        \path [line] (1.east) to[out=0, in=90] (2.60);
        \path [line] (2.-60) to[out=-90, in=0] (3.east);
        \path [line] (3.west) to[out=180, in=-90] (4.-120);
        \path [line] (4.120) to[out=90, in=180] (1.west);
    \end{tikzpicture}}
    \caption{The PIC loop, which is repeated every time-step after the initialization.}
    \label{pic_loop}
\end{figure}

Let $\vec{r}_i$ be the position of macro-particle $i$ with mass $m_i$ and charge $q_i$, $\vec{p}_i=m\gamma\vec{v}_i$ its relativistic momentum, $\vec{v}_i$ its velocity, and $\gamma_i=1/\sqrt{1-|\vec{v}_i|^2/c^2}$ the Lorentz factor, where $c$ is the speed of light. The equations of motion are given by the Newton-Lorentz equations:
\begin{align*}
    \frac{d\vec{r}_i}{dt} & = \frac{\vec{p}_i}{m_i\gamma_i} \\
    \frac{d\vec{p}_i}{dt} & = q_i(\vec{E}_i + \frac{\vec{p}_i}{m_i\gamma_i}\times \vec{B}_i).
\end{align*}
The above are integrated using a Leapfrog algorithm \cite{verboncoeur_particle_2005} in the particle push phase to evolve all of the macro-particles ($i=1, ..., N_p$ where $N_p$ is the number of macro-particles) to the next timestep. The fields $\vec{E}_i$ and $\vec{B}_i$ are the sum of the self-fields of the particles and any external fields that may be applied, $\vec{E}_{ext}$ and $\vec{B}_{ext}$. In the electrostatic case, the magnetic self-field is zero in the non-relativistic regime. In the case of relativistic particles, a magnetic self-field appears due to the moving charges.

Let $\vec{x}\in\mathbb{R}$ represent a position in the computational mesh, for fields (not the same as the macro-particle positions). At each time-step, the electric field $\vec{E}(\vec{x})$ is obtained from
\begin{equation} \label{E-field}
    \vec{E}(\vec{x}) = -\vec{\nabla} \phi(\vec{x}),
\end{equation}
where the potential $\phi(\vec{x})$ is computed from the charge density $\rho(\vec{x})$ using \cref{poisson_eq}. After calculating the electric field $\vec{E}(\vec{x})$ on the mesh, it is interpolated as $\vec{E}_i$ to each macro-particle located at $\vec{r}_i$ (gather phase).

As mentioned in the previous section, the Poisson equation in the \textbf{solve} phase of the PIC loop is written as a convolution and solved in Fourier space. In the case of free-space boundaries, one needs to to use the Hockney-Eastwood or the Vico-Greengard method as explained before.

\subsection{Penning trap as a physics example}

\begin{figure}[ht]
\centering
\begin{subfigure}{.3\textwidth}
  \centering
  \includegraphics[width=\linewidth]{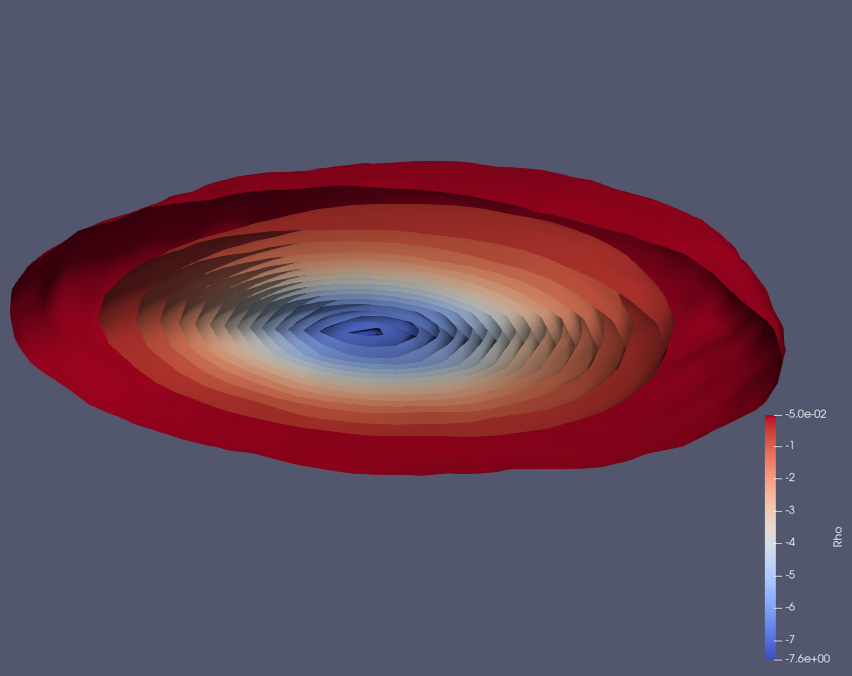}
  \caption{Time = 0}
\end{subfigure}\hfill
\begin{subfigure}{.3\textwidth}
  \centering
  \includegraphics[width=\linewidth]{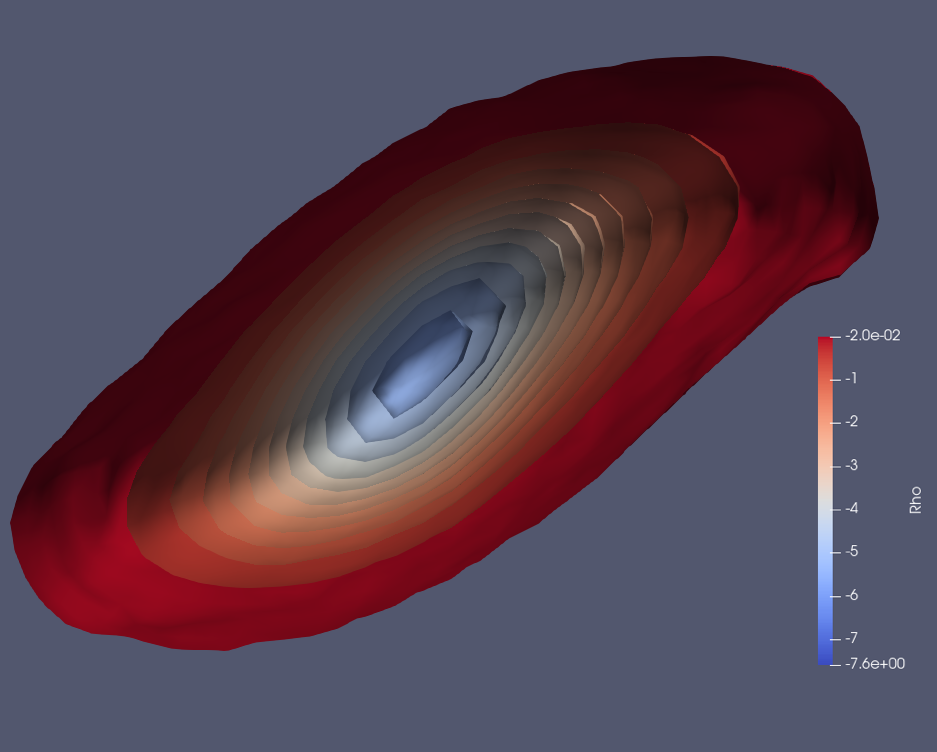}
  \caption{Time = 1.95}
\end{subfigure}\hfill
\begin{subfigure}{.3\textwidth}
  \centering
  \includegraphics[width=\linewidth]{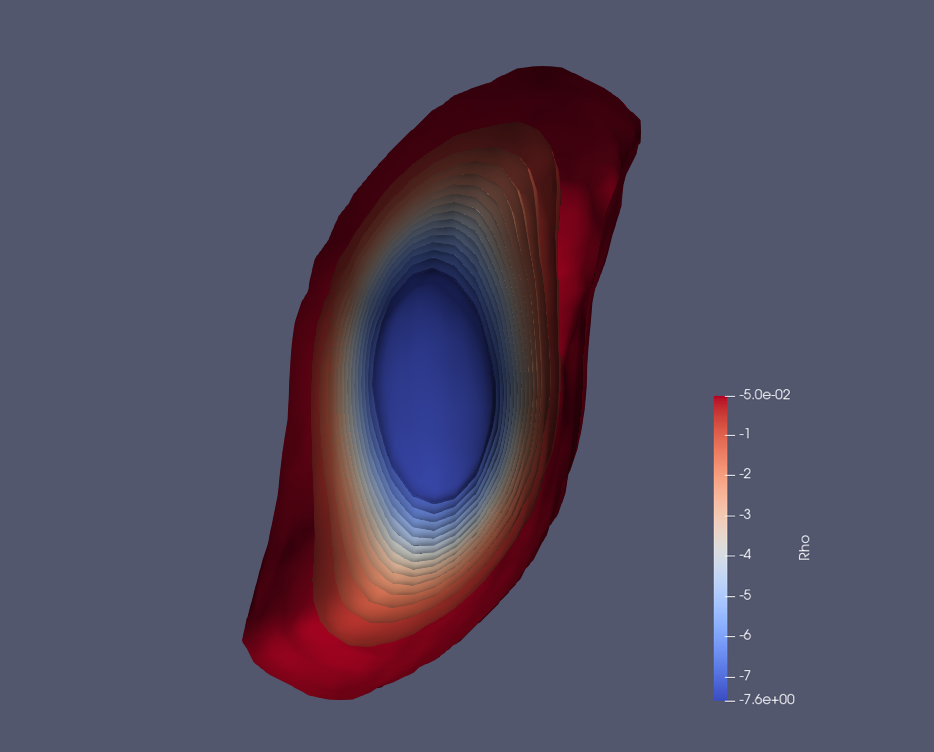}
  \caption{Time = 3.91}
\end{subfigure}
\begin{subfigure}{.3\textwidth}
  \centering
  \includegraphics[width=\linewidth]{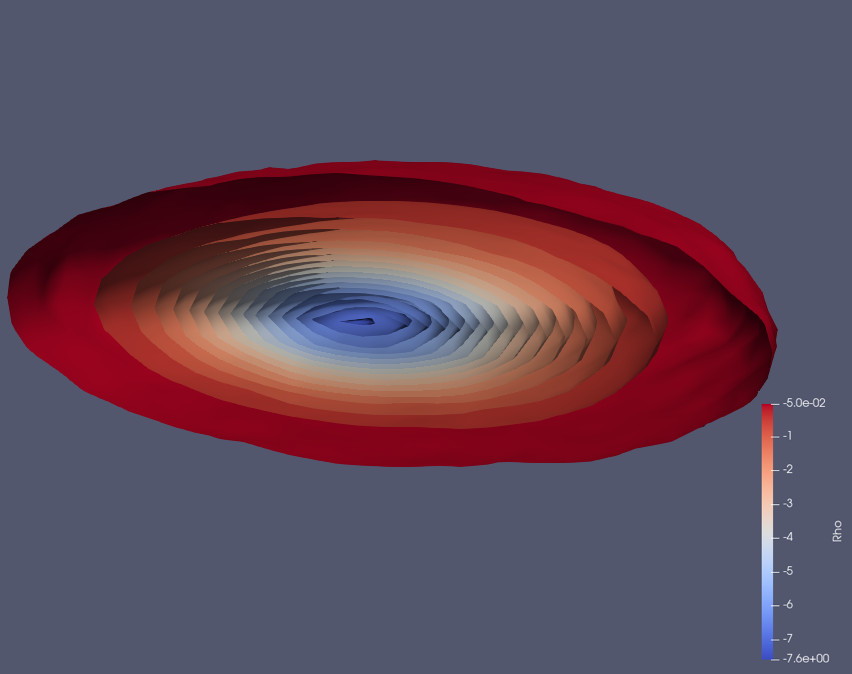}
  \caption{Time = 0}
\end{subfigure}\hfill
\begin{subfigure}{.3\textwidth}
  \centering
  \includegraphics[width=\linewidth]{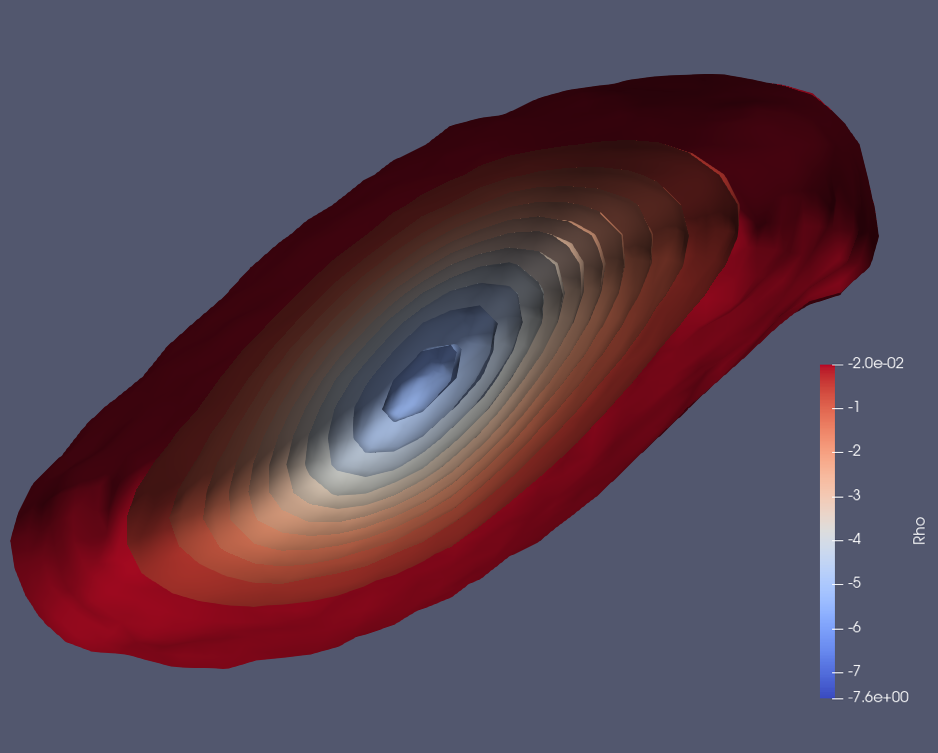}
  \caption{Time = 1.95}
\end{subfigure}\hfill
\begin{subfigure}{.3\textwidth}
  \centering
  \includegraphics[width=\linewidth]{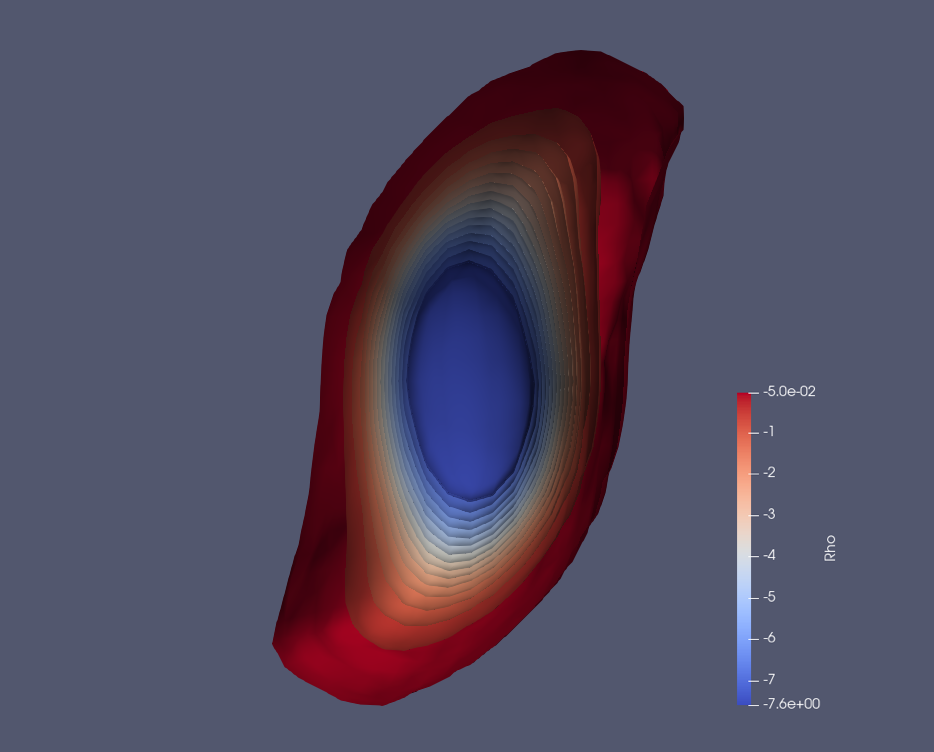}
  \caption{Time = 3.91}
\end{subfigure}
\caption{Evolution of electron bunch dynamics in a 3D charge-neutral Penning trap, shown as snapshots in time during the simulation. The first and second rows correspond to using the Hockney-Eastwood and Vico-Greengard solvers, respectively.}
\label{penning}
\end{figure}

For the Penning trap, we follow the same set-up as in \cite{muralikrishnan_scaling_2024}, with open boundary conditions. We do a qualitative comparison between using the Hockney-Eastwood solver and the Vico-Greengard solver by looking at the evolution of the electron charge density, as shown in \cref{penning}. Both solvers produce the same expected behaviour with similar runtimes.

Identifying the regimes or test cases where the high accuracy of the Vico-Greengard solver is beneficial is beyond the scope of the current work and will be carried as a part of future work. 

\section{Memory Performance Analysis Results} \label{performance_analytics}

Here we provide the full profiling measurements of the heFFTe benchmarks, which we performed using their own benchmark output measurements and the mpiP tool \cite{llnlmpip_llnlmpip_2006}, which provides statistics on MPI calls in the program. Tables \ref{tablec1}, \ref{tablec2}, and \ref{tablec3} contain the data for the Perlmutter system, Alps, and Lumi, respectively.

\subsection{Perlmutter}

\resizebox{\columnwidth}{!}{
\begin{tabular}{rrrrrr}
\toprule
Problem Size & GPU Count & Performance (GFLOPs/s) & Time (s) & Bandwidth (GB/s) & FLOPs/Byte \\
\midrule
$1024^3$ & 16 & 560.271 & 0.287470 & 1055.017915 & 0.531054 \\
$1024^3$ & 32 & 863.627 & 0.186494 & 1712.769312 & 0.504228 \\
$1024^3$ & 64 & 1280.930 & 0.125738 & 2860.735816 & 0.447762 \\
$2048^3$ & 64 & 1597.360 & 0.887303 & 3240.413929 & 0.492949 \\
$2048^3$ & 128 & 2787.450 & 0.508471 & 5769.331978 & 0.483150 \\
$2048^3$ & 256 & 4781.280 & 0.296435 & 10081.383777 & 0.474268 \\
\bottomrule
\end{tabular}
\label{tablec1}
}

\subsection{ALPS}

\resizebox{\columnwidth}{!}{
\begin{tabular}{rrrrrr}
\toprule
Problem Size & GPU Count & Performance (GFLOPs/s) & Time (s) & Bandwidth (GB/s) & FLOPs/Byte \\
\midrule
$1024^3$ & 16 & 913.994 & 0.176217 & 1721.037130 & 0.531072 \\
$1024^3$ & 32 & 1277.160 & 0.126109 & 2533.023020 & 0.504204 \\
$1024^3$ & 64 & 1775.930 & 0.090691 & 3966.488517 & 0.447734 \\
$2048^3$ & 64 & 2450.920 & 0.578288 & 4972.446255 & 0.492900 \\
$2048^3$ & 128 & 4158.120 & 0.340860 & 8606.160887 & 0.483156 \\
$2048^3$ & 256 & 8044.070 & 0.176197 & 16961.026578 & 0.474268 \\
\bottomrule
\end{tabular}
\label{tablec2}
}

\subsection{Lumi}

\resizebox{\columnwidth}{!}{
\begin{tabular}{rrrrrr}
\toprule
Problem Size & GPU Count & Performance (GFLOPs/s) & Time (s) & Bandwidth (GB/s) & FLOPs/Byte \\
\midrule
$1024^3$ & 16 & 125.161 & 1.286830 & 235.684589 & 0.531053 \\
$1024^3$ & 32 & 231.492 & 0.695754 & 459.089276 & 0.504242 \\
$1024^3$ & 64 & 384.548 & 0.418833 & 858.856871 & 0.447744 \\
$2048^3$ & 64 & 432.246 & 3.279010 & 876.932062 & 0.492907 \\
$2048^3$ & 128 & 826.134 & 1.715630 & 1709.954944 & 0.483132 \\
$2048^3$ & 256 & 1553.110 & 0.912579 & 3274.775115 & 0.474265 \\
\bottomrule
\end{tabular}
\label{tablec3}
}

\clearpage





% \end{thebibliography}
\end{document}